\documentclass[12pt]{article}
\usepackage{color}
\usepackage{amssymb,amsmath,comment}
\usepackage{graphicx}
\usepackage{braket}
\usepackage{epsfig}
\usepackage{here}
\graphicspath{{./Figures/}}
\newcommand{\ba}{\begin{align}}
\newcommand{\ea}{\end{align}}

\def\nn{\nonumber}

\def\bea{\begin{eqnarray}}
\def\eea{\end{eqnarray}}
 \makeatletter
\def\alt{\mathrel{\mathpalette\gl@align<}}
\def\agt{\mathrel{\mathpalette\gl@align>}}
\def\gl@align#1#2{\lower.6ex\vbox{\baselineskip\z@skip\lineskip\z@
\ialign{$\m@th#1\hfil##\hfil$\crcr#2\crcr\sim\crcr}}} \makeatother
\renewcommand{\thefootnote}{\fnsymbol{footnote}}
\begin{document}
\begin{flushright}
\end{flushright}
\vspace*{1.0cm}

\begin{center}
\baselineskip 20pt 
{\Large\bf 
On $\theta_{23}$ Octant Measurement in $3+1$ Neutrino Oscillations \\
in T2HKK
}
\vspace{1cm}

{\large 
Naoyuki Haba,
\ Yukihiro Mimura
\ and \ Toshifumi Yamada
} \vspace{.5cm}

{\baselineskip 20pt \it
Institute of Science and Engineering, Shimane University, Matsue 690-8504, Japan
}

\vspace{.5cm}

\vspace{1.5cm} {\bf Abstract} \end{center}

It has been pointed out that the mixing of an eV-scale sterile neutrino with active flavors can
 lead to loss of sensitivity to the $\theta_{23}$ octant (sign of $\sin^2\theta_{23}-1/2$) in long baseline experiments,
 because the main oscillation probability $P_0=4\sin^2\theta_{23}\sin^2\theta_{13}\sin^2\Delta_{31}$
 can be degenerate with the sum of the interferences with the solar oscillation amplitude and an active-sterile oscillation amplitude
 in both neutrino and antineutrino oscillations, depending on CP phases.
In this paper, we show that the above degeneracy is resolved by measuring the same beam at different baseline lengths.
We demonstrate that Tokai-to-Hyper-Kamiokande-to-Korea (T2HKK) experiment 
 (one 187~kton fiducial volume water Cerenkov detector is placed at Kamioka, $L=295$~km, and another detector is put in Korea, $L\sim1000$~km)
 exhibits a better sensitivity to the $\theta_{23}$ octant in those parameter regions where the experiment with two detectors at Kamioka is insensitive to it.
Therefore, if a hint of sterile-active mixings is discovered in short baseline experiments, T2HKK is a better option than the plan of placing two detectors at Kamioka.
We also consider an alternative case where one detector is placed at Kamioka and a different detector is at Oki Islands, $L=653$~km,
 and show that this configuration also leads to a better sensitivity to the $\theta_{23}$ octant.

\thispagestyle{empty}

%\bigskip
\newpage
\renewcommand{\thefootnote}{\arabic{footnote}}
%\addtocounter{page}{-1}
\setcounter{footnote}{0}
%%%%%%%%%%%%%%%%%%%%%%%%%%
%\baselineskip 36pt
% Main body
%%%%%%%%%%%%%%%%%%%%%%%%%%
\baselineskip 18pt
%%%%%%%%%%%%%%%%%%%%%%%%%%
\section{Introduction}

The octant of neutrino mixing angle $\theta_{23}$, namely, the sign of $\sin^2\theta_{23}-1/2$,
 has profound implications for SO(10) grand unification theory with renormalizable Yukawa couplings
 (see, e.g., Refs.~\cite{BhupalDev:2012nm,Fukuyama:2015kra,Fukuyama:2016vgi}).
Since $\nu_e$ and $\nu_\mu$ disappearance channels are almost solely sensitive to $\sin^2(2\theta_{23})$,
 the best way to determine the octant is to use $\nu_\mu\to\nu_e$ channel in long baseline experiments.
For the standard 3-flavor oscillations, the octant measurement has been investigated in 
 Refs.~\cite{Barger:2001yr,Hagiwara:2006nn,Meloni:2008bd,Agarwalla:2013ju,Chatterjee:2013qus,Minakata:2013hgk,Agarwalla:2013hma,Choubey:2013xqa,Bora:2014zwa,Das:2014fja,Ghosh:2015ena,Nath:2015kjg}.

It has been pointed out that if an eV-scale light sterile neutrino mixes with active flavors,
 it can considerably deteriorate the sensitivity of long baseline experiments to the $\theta_{23}$ octant~\cite{Agarwalla:2016xlg}
 \footnote{
Recent studies on the impact of sterile-active mixings on long baseline experiments are 
 Refs.~\cite{Klop:2014ima,Agarwalla:2016mrc,Choubey:2016fpi,Ghosh:2017atj,Choubey:2017cba,Choubey:2017ppj,Agarwalla:2018,Choubey:2018kqq}.
}.
This is based on the following observation:
In the standard 3-flavor oscillations,
 the part of the $\nu_\mu\to\nu_e$ oscillation probability sensitive to the $\theta_{23}$ octant,
 $P_0=4\sin^2\theta_{23}\sin^2\theta_{13}\sin^2\Delta_{31}$ $(\Delta_{31}\equiv\Delta m_{31}^2L/(4E)$),
 is possibly mimicked by the interference of the solar and atmospheric oscillation amplitudes,
 $P_1\simeq4\sin\theta_{13}\sin\theta_{12}\cos\theta_{12}(\Delta m^2_{21}/\Delta m^2_{31}) \Delta_{31}\sin\Delta_{31}\cos(\Delta_{31}\pm\phi_{13})$,
 in either neutrino oscillations or antineutrino oscillations depending on the CP phase $\phi_{13}$, 
 but not in both oscillations.
Hence, a combination of neutrino- and antineutrino-focusing operations resolves the above degeneracy.
In the $3+1$ oscillations, however, the interference of atmospheric and active-sterile oscillation amplitudes
 generates a new term, $P_2\simeq4\sin\theta_{14}\sin\theta_{24}\sin\theta_{13}\sin\theta_{23}
 \sin\Delta_{31}\sin(\Delta_{31}\pm\phi_{13}\mp\phi_{14})$.
The sum $P_1+P_2$ can mimic $P_0$ in both neutrino and antineutrino oscillations for some values of $\phi_{13}$ and $\phi_{14}$, leading to loss of sensitivity to the $\theta_{23}$ octant.

In this paper, we pursue the possibility that
 a combination of neutrino- and antineutrino-focusing operations \textit{and}
 measurements of the same beam at different baseline lengths
 resolves the degeneracy of $P_0$ and $P_1+P_2$, thereby resurrecting the sensitivity to the $\theta_{23}$ octant 
 in the $3+1$ oscillations.
For concreteness, we concentrate on Tokai-to-Hyper-Kamiokande-to-Korea (T2HKK) experiment~\cite{Abe:2016ero}
 (for early proposals, see 
 Refs.~\cite{Hagiwara:2004iq,Ishitsuka:2005qi,Hagiwara:2005pe,Hagiwara:2006vn,Kajita:2006bt,Hagiwara:2006nn,Huber:2007em,Hagiwara:2009bb,Dufour:2010vr}),
 where one 187~kton fiducial volume water Cerenkov detector is placed at Kamioka ($L=295$~km) and another 187~kton detector is in Korea ($L\sim1000$~km),
 which is a proposed extension of Tokai-to-Hyper-Kamiokande (T2HK) experiment~\cite{Abe:2018uyc}.
We conduct a comparative study of T2HKK experiment with the plan of placing two 187~kton detectors at Kamioka.
We will demonstrate that for some values of CP phases $\phi_{13}$ and $\phi_{14}$,
 the sensitivity to the $\theta_{23}$ octant is lost in the latter experiment
 while the sensitivity is maintained in T2HKK, in spite of smaller statistics of T2HKK
 (as the baseline to Korea is longer than that to Kamioka).
We further consider an alternative plan of placing a modest detector 
 at Oki Islands~\cite{Badertscher:2008bp,Hagiwara:2012mg,Hagiwara:2016qtb,Abe:2017jit} ($L=653$~km)
  in addition to one 187~kton detector at Kamioka, and study how the sensitivity to the $\theta_{23}$ octant changes in this case.

Previously, the sensitivity of T2HK and T2HKK to the $\theta_{23}$ octant in the presence of mixings of an eV-scale sterile neutrino
 has been studied in Ref.~\cite{Choubey:2017cba}.
Our study differs from it in that we consider both cases with $\theta_{34}=0$ and $\theta_{34}\neq0$, and further separately investigate the dependence of the sensitivity on $\theta_{34}$ and that on $\theta_{24}$
\footnote{
Since the impact of sterile-active mixings on the $\theta_{23}$ octant measurement 
 appears only in the combinations $\sin\theta_{14}\sin\theta_{24}$ and 
 $\sin\theta_{14}\sin\theta_{34}$, we do not need to vary $\theta_{14}$ in addition to $\theta_{34}$ and $\theta_{24}$.
}.
In contrast, Ref.~\cite{Choubey:2017cba} only assumes substantially large values for $\theta_{34}$ and varies $\theta_{34},\theta_{14},\theta_{24}$ only simultaneously.
As we show in the main text, non-zero values of $\theta_{34}$ tend to increase the sensitivity to the $\theta_{23}$ octant,
 and hence it is important to scrutinize the case with $\theta_{34}=0$, which is the `worst situation' for the $\theta_{23}$ octant measurement.
Also, since the sensitivity tends to increase with $\theta_{34}$ and decrease with $\theta_{24}$, the dependences on $\theta_{34}$ and $\theta_{24}$ must be analyzed separately.

This paper is organized as follows:
In Section~2, we write the $\nu_\mu\to\nu_e$ oscillation probability in the $3+1$ oscillations, 
spot the terms that can lead to loss of sensitivity to $\theta_{23}$ octant, and qualitatively state that this problem is mitigated in T2HKK experiment.
In Section~3, we confirm the above qualitative statement through a numerical analysis.
Section~4 summarizes the paper.

%%%%%%%%%%%%%%
\section{$3+1$ Oscillation Probability}

We postulate the presence of one isospin-singlet neutrino, $\nu_s$, that mixes with the three active flavors $(\nu_e,\nu_\mu,\nu_\tau)$ and
 yields four mass eigenstates $(\nu_1,\nu_2,\nu_3,\nu_4)$.
We also assume that the mass of $\nu_4$ is around 1~eV and thus $\nu_4$ participates in oscillations in a J-PARC beam.

We parametrize the mixing of $(\nu_e,\nu_\mu,\nu_\tau,\nu_s)$ as follows:
\bea
\vert\nu_\alpha\rangle=\sum_{k=1}^4(U)_{\alpha k}\vert \nu_k\rangle \ \ \ \ \ (\alpha=e,\mu,\tau,s),
\eea
 with $U$ being a 4$\times$4 unitary matrix defined as
\bea
&&U=U_{34}U_{24}U_{14}U_{23}U_{13}U_{12},
\nn\\
&&(U_{ab})_{ij}=\delta_{ia}\delta_{ja}\cos\theta_{ab}+\delta_{ib}\delta_{jb}\cos\theta_{ab}
+\delta_{ia}\delta_{jb}e^{-i\phi_{ab}}\sin\theta_{ab}-\delta_{ib}\delta_{ja}e^{i\phi_{ab}}\sin\theta_{ab}.
\nn\\
&& \ \ \ \ \ \ \ \ \ \ \ \ \ \ \ \ \ \ \ \ \ \ \ \ \ \ \ \ \ \ \ \ \ \ \ \ \ \ (a,b=1,2,3,4;\,a<b)
\nn
\eea
$U_{23}U_{13}U_{12}$ corresponds to Pontecorvo-Maki-Nakagawa-Sakata matrix~\cite{mns,p}.
Henceforth, we fix the phase convention such that $\phi_{12}=\phi_{23}=\phi_{24}=0$.
Also, we take $\theta_{14}\geq0,\,\theta_{24}\geq0,\,\theta_{34}\geq0$.

The probability of $\nu_\mu\to\nu_e$ oscillation, which is crucial for the $\theta_{23}$ octant measurement, is given by
\bea
&&P(\nu_\mu\to\nu_e)=\left\vert\left[\exp\left(-i\int_0^L{\rm d}x\ H(x)\right)\right]_{12}\right\vert^2,
 \label{prob}\\
&& \ \ \ \ \ \ \ H(x)= \frac{1}{2E}U   \begin{pmatrix} % or pmatrix or bmatrix or Bmatrix or ...
      0 & 0 & 0 & 0 \\
      0 & \Delta m_{21}^2 & 0 & 0 \\
       0 & 0 & \Delta m_{31}^2 & 0 \\
        0  & 0 & 0 & \Delta m_{41}^2 \\
   \end{pmatrix} U^\dagger
+   \begin{pmatrix} % or pmatrix or bmatrix or Bmatrix or ...
      V_{cc}+V_{nc} & 0 & 0 & 0 \\
      0 & V_{nc} & 0 & 0 \\
      0 & 0 & V_{nc} & 0 \\
      0 & 0 & 0 &0 \\
   \end{pmatrix},\nn
\eea
 where $L$ is the baseline length, $E$ is the neutrino energy, and $\Delta m_{i1}^2=m_i^2-m_1^2$.
$V_{cc}$ and $V_{nc}$ denote
 the potentials generated by charged and neutral current interactions, respectively,
 which are evaluated as $V_{cc}=-2V_{nc}\simeq0.193\times10^{-3}(\rho/({\rm g}/{\rm cm}^3))/{\rm km}$ with
 $\rho$ being the matter density.
The antineutrino oscillation probability $P(\bar{\nu}_\mu\to\bar{\nu}_e)$ is obtained by
 flipping the signs of $\phi_{13},\,\phi_{14},\,\phi_{34},\,V_{cc},\,V_{nc}$.

To gain physical insight, we expand Eq.~(\ref{prob}) in the leading order of 
 $\Delta m_{21}^2L/E,~\theta_{14},~\theta_{24},~\theta_{34}$ and the matter effect (approximated to be uniform),
 and further take an average over $\Delta m_{41}^2L/(4E)$ oscillations.
The $\nu_\mu\to\nu_e$ and $\bar{\nu}_\mu\to\bar{\nu}_e$ oscillation probabilities are then written as
\begin{align}
P(\stackrel{{\mbox {\tiny(}}-{\mbox {\tiny)}}}{\nu}_\mu\to\stackrel{{\mbox {\tiny(}}-{\mbox {\tiny)}}}{\nu}_e)\simeq&\
4\sin^2\theta_{13}\sin^2\theta_{23}\sin^2\left(\frac{\Delta m_{31}^2L}{4E}\right)
\nn\\
&+4\sin\theta_{13}\sin\theta_{12}\cos\theta_{12}\sin(2\theta_{23})\frac{\Delta m_{21}^2L}{4E}
\sin\left(\frac{\Delta m_{31}^2L}{4E}\right)\cos\left(\frac{\Delta m_{31}^2L}{4E}\pm\phi_{13}\right)
\nn\\
&+4\sin\theta_{14}\sin\theta_{24}\sin\theta_{13}\sin\theta_{23}\sin\left(\frac{\Delta m_{31}^2L}{4E}\right)\sin\left(\frac{\Delta m_{31}^2L}{4E}\pm\phi_{13}\mp\phi_{14}\right)
\nn\\
&\pm4L\,V_{cc}\sin^2\theta_{13}\sin^2\theta_{23}\left\{\frac{4E}{\Delta m_{31}^2L}\sin^2\left(\frac{\Delta m_{31}^2L}{4E}\right)
-\frac{1}{2}\sin\left(\frac{\Delta m_{31}^2L}{2E}\right)\right\}
\nn\\
&\pm\sin\theta_{13}\sin\theta_{23}\sin(2\theta_{23})\sin\theta_{14}\sin\theta_{34}\,L\,V_{nc}\cdot
F\left(\frac{L\Delta m_{31}^2}{4E},\pm(\phi_{13}-\phi_{14}+\phi_{34})\right)
\nn\\
=&\ 4\sin^2\theta_{13}\sin^2\theta_{23}\sin^2\left(\frac{\Delta m_{31}^2L}{4E}\right)
\label{main}\\
&+\left\{\mp A\sin\phi_{13}+B\cos(\phi_{13}-\phi_{14})\right\}\ \sin^2\left(\frac{\Delta m_{31}^2L}{4E}\right)
\label{sinsq}\\
&+\frac{1}{2}\left\{A\cos\phi_{13}\pm B\sin(\phi_{13}-\phi_{14})\right\}\ \sin\left(\frac{\Delta m_{31}^2L}{2E}\right)
\label{sin2}\\
&\pm4L\,V_{cc}\sin^2\theta_{13}\sin^2\theta_{23}\left\{\frac{4E}{\Delta m_{31}^2L}\sin^2\left(\frac{\Delta m_{31}^2L}{4E}\right)
-\frac{1}{2}\sin\left(\frac{\Delta m_{31}^2L}{2E}\right)\right\}
\label{matter}\\
&\pm\sin\theta_{13}\sin\theta_{23}\sin(2\theta_{23})\sin\theta_{14}\sin\theta_{34}\,L\,V_{nc}\cdot
F\left(\frac{L\Delta m_{31}^2}{4E},\pm(\phi_{13}-\phi_{14}+\phi_{34})\right),
\label{matter2}
\end{align}
 where
\begin{align}
&A=4\sin\theta_{13}\sin\theta_{12}\cos\theta_{12}\sin(2\theta_{23})\frac{\Delta m_{21}^2L}{4E},
\nn\\
&B=4\sin\theta_{14}\sin\theta_{24}\sin\theta_{13}\sin\theta_{23},
\end{align}
 and the upper signs in $\pm,\,\mp$ refer to the neutrino oscillation, and the lower signs to the antineutrino oscillation.

First, we find that in the absence of sterile-active mixings, we have $B=0$ and the impact of $A$ on the $\theta_{23}$ octant measurement
 can be reduced by a combination of neutrino- and antineutrino-focusing beams, 
 since $A$ term in Eq.~(\ref{sinsq}) changes sign for neutrino and antineutrino.

Second, we observe that in the presence of sterile-active mixings (i.e. $B\neq0$),
 a combination of neutrino- and antineutrino-focusing beams \textit{and}
 two different baseline lengths revives the sensitivity to the $\theta_{23}$ octant,
 because a different oscillating function, $\sin(\Delta m_{31}^2L/(2E))$, enters Eq.~(\ref{sin2}).
In the rest of the paper, we will confirm this qualitative observation through simulations of T2HK, T2HKK, and T2HK+Oki experiments.

The term Eq.~(\ref{matter}) represents the ordinary matter effect that is present in the standard oscillations, and does not affect the measurement of the $\theta_{23}$ octant.

The last term Eq.~(\ref{matter2}) represents a synergy of the matter effect and sterile-active mixings.
Here, $F$ is a complicated function of $L\Delta m_{31}^2/(4E)$ and $\pm(\phi_{13}-\phi_{14}+\phi_{34})$, and is expected to be $O(1)$.
Since $\theta_{34}$ is less constrained than $\theta_{24}$, this term can lead to interesting phenomenology 
 in some long baseline experiments with large matter effect.
We will study the impact of the term Eq.~(\ref{matter2}) numerically in the next section.
\\

\section{Numerical Study}

We simulate the propagation of neutrinos and antineutrinos, their $3+1$ oscillations, and their measurements at Kamioka, Oki and Korea.
We calculate $\nu_\mu\to\nu_e$, $\bar{\nu}_\mu\to\bar{\nu}_e$, 
 $\nu_\mu\to\nu_\mu$, $\bar{\nu}_\mu\to\bar{\nu}_\mu$ oscillation probabilities
 using Eq.~(\ref{prob}) without approximation.
\\

\subsection{Oscillation Parameters}

We make a simplifying assumption that $\theta_{12},\,\theta_{13},\,\Delta m^2_{21},\,\vert\Delta m^2_{32}\vert$
 and the true mass hierarchy have been measured precisely prior to a $\theta_{23}$ measurement.
Accordingly, we fix the values of $\theta_{12},\,\theta_{13},\,\Delta m^2_{21},\,\vert\Delta m^2_{32}\vert$
 at their Particle Data Group values~\cite{Tanabashi:2018oca} in the simulation, and fit the simulation results with the true values.
Likewise, we perform the simulation for the normal hierarchy case ($\Delta m^2_{32}>0$) and fit the results by assuming the normal hierarchy, and do analogously for the inverted hierarchy case ($\Delta m^2_{32}<0$).

For sterile-active mixing parameters,
 we consider a situation where sterile-active mixings have already been discovered and 
 $\Delta m_{41}^2,\,\theta_{14},\,\theta_{24},\,\theta_{34}$ have been measured precisely in short baseline experiments.

The full analysis is multi-dimensional, depending on benchmark values of $\theta_{14},\,\theta_{24},\,\theta_{34}$
 as well as unknown values of $\theta_{23}$ and
 the three CP phases $\phi_{13},\,\phi_{14},\,\phi_{34}$.
Such a complicated analysis does not provide clear physical insight.
Therefore, first we restrict our study to the case with $\theta_{34}=0$, and assume the largest values for $\theta_{14},\,\theta_{24}$
 that are consistent with the current experimental bounds in order to {\it assess the maximal impact of $\theta_{14},\,\theta_{24}$ mixings on the
 $\theta_{23}$ octant measurement}.
The reason for taking $\theta_{34}=0$ is that 
 the $\theta_{34}$ mixing affects the octant measurement only via the matter effect
 and so its impact should be studied separately.
Here we fix $\Delta m_{41}^2$ at a value around $\sim1$~eV$^2$, 
 since the results do not depend on the precise value of $\Delta m_{41}^2$.
In contrast, we consider various values for $\phi_{13},\,\phi_{14}$ because the impact of $\theta_{14},\,\theta_{24}$ mixings
 on the $\theta_{23}$ octant measurement depends keenly on $\phi_{13},\,\phi_{14}$.
We take two benchmark values for $\theta_{23}$, one in the lower octant and the other in the higher octant.
The first benchmark parameter set, which is motivated by the above argument, is presented in Table~\ref{physpara}.
\begin{table}[H]
\begin{center}
  \caption{Parameters in the first benchmark.}
  \begin{tabular}{|c||c|} \hline
    parameter & value in our simulation \\ \hline
    $\sin^2\theta_{12}$ & 0.304 \\
    $\sin^2\theta_{13}$ & 0.0219 \\
    $\sin^2\theta_{23}$ & 0.44 \ or \ 0.60 \\
    $\phi_{13}$ (rad) & $[0,\,2\pi]$ \\ 
    $\Delta m^2_{21}$ & $7.53\times10^{-5}$~eV$^2$  \\
    $\Delta m^2_{32}$ (normal hierarchy)   & $2.51\times10^{-3}$~eV$^2$  \\
    $\Delta m^2_{32}$ (inverted hierarchy) & $-2.56\times10^{-3}$~eV$^2$ \\ \hline
    $\sin^2\theta_{14}$ & 0.04\\
    $\sin^2\theta_{24}$ & 0.006\\
    $\sin^2\theta_{34}$ & 0 \\
    $\phi_{14}$ (rad) & $[0,\,2\pi]$ \\
    $\phi_{34}$ (rad) & unphysical \\
    $\Delta m^2_{41}$ & 3~eV$^2$ \\ \hline
  \end{tabular}
  \label{physpara}
  \end{center}
\end{table}
In Table~\ref{physpara}, the values of $\sin^2\theta_{14}$ and $\Delta m^2_{41}$
 are marginally compatible with the 90\% CL bound obtained by a reanalysis~\cite{Adamson:2016jku}
 of Bugey-3~\cite{Declais:1994su} experiment,
 and the values of $\sin^2\theta_{24}$ and $\Delta m^2_{41}$
 are marginally consistent with the 90\% CL bound of the
 MINOS and MINOS+ results~\cite{Adamson:2017uda}
 and are outside the 90\% CL bound from the IceCube~\cite{TheIceCube:2016oqi}.
The asymmetric benchmark values of $\sin^2\theta_{23}$ are motivated by the $3\sigma$ range reported by NuFIT~4.1~\cite{Esteban:2018azc,nufit}, which is tilted towards the higher octant region.
Note that $\phi_{34}$ is an unphysical phase when $\theta_{34}=0$.
\\

After the analysis of the first benchmark Table~\ref{physpara},
 we vary $\sin^2\theta_{34},\,\phi_{34}$ and study the impact of a synergy of the matter effect and sterile-active mixings
 described by Eq.~(\ref{matter2}).
Here we fix $\phi_{13},\,\phi_{14}$ at the values for which sterile-active mixings have most afflicted
 the $\theta_{23}$ octant measurement in the first benchmark.
The second benchmark parameter set, which is motivated by the above argument, is presented in Table~\ref{physpara2}.
\begin{table}[H]
\begin{center}
  \caption{Parameters in the second benchmark.
  The values of $\phi_{13},\phi_{14}$ are such that sterile-active mixings most afflict the $\theta_{23}$ octant measurement
  in the first benchmark Table~\ref{physpara}, where $\sin^2\theta_{34}=0$.}
  \begin{tabular}{|c||c|} \hline
    parameter & value in our simulation \\ \hline
    $\sin^2\theta_{12}$ & 0.304 \\
    $\sin^2\theta_{13}$ & 0.0219 \\
    $\sin^2\theta_{23}$ & 0.44 \ or \ 0.60 \\
    $\phi_{13}$ (rad) & 0 when $\sin^2\theta_{23}=0.44$, \ \ \ $\pi$ when $\sin^2\theta_{23}=0.60$ \\ 
    $\Delta m^2_{21}$ & $7.53\times10^{-5}$~eV$^2$  \\
    $\Delta m^2_{32}$ (normal hierarchy)   & $2.51\times10^{-3}$~eV$^2$  \\
    $\Delta m^2_{32}$ (inverted hierarchy) & $-2.56\times10^{-3}$~eV$^2$ \\ \hline
    $\sin^2\theta_{14}$ & 0.04\\
    $\sin^2\theta_{24}$ & 0.006\\
    $\sin^2\theta_{34}$ & $[0,\,0.18]$ \\
    $\phi_{14}$ (rad) & 0 \\
    $\phi_{34}$ (rad) & $[0,\,2\pi]$ \\
    $\Delta m^2_{41}$ & 3~eV$^2$ \\ \hline
  \end{tabular}
  \label{physpara2}
  \end{center}
\end{table}
In Table~\ref{physpara2}, the range of $\sin^2\theta_{34}$ is chosen to
 satisfy the 90\% CL bound on the sterile-tau neutrino mixing reported from Super-Kamiokande atmospheric neutrino measurement~\cite{Abe:2014gda}.
\\

We will find in Section~\ref{results} that non-zero values of $\sin^2\theta_{34}$ tend to mitigate the impact of sterile-active mixings on the $\theta_{23}$ octant measurement, i.e. one is more likely to obtain the correct octant when the $\theta_{34}$ mixing is present.
Therefore, it is important to study how the impact of sterile-active mixings changes with $\theta_{14},\,\theta_{24}$
 in the `worst scenario' for the octant measurement where $\theta_{34}=0$.
To this end, we vary $\theta_{24}$ while fixing $\theta_{34}=0$.
Here we fix $\sin^2\theta_{14}=0.04$ because, as seen in Eqs.~(\ref{main})-(\ref{matter2}), the effects of sterile-active mixings
 on the $\theta_{23}$ octant measurement
 appear only in the combination $\sin\theta_{14}\sin\theta_{24}$ when $\theta_{34}=0$,
 and thus varying $\sin^2\theta_{14}$ is equivalent to varying $\sin^2\theta_{24}$.
Here we take a specific combination of values of $\phi_{13},\,\phi_{14}$ for which sterile-active mixings have most afflicted
 the $\theta_{23}$ octant measurement in the first benchmark.
The third benchmark parameter set, which is motivated by the above argument, is presented in Table~\ref{physpara3}.
\begin{table}[H]
\begin{center}
  \caption{Parameters in the third benchmark.
  The values of $\phi_{13},\phi_{14}$ are such that sterile-active mixings most afflict the $\theta_{23}$ octant measurement
  in the first benchmark Table~\ref{physpara}, where we fix $\sin^2\theta_{24}=0.006$.}
  \begin{tabular}{|c||c|} \hline
    parameter & value in our simulation \\ \hline
    $\sin^2\theta_{12}$ & 0.304 \\
    $\sin^2\theta_{13}$ & 0.0219 \\
    $\sin^2\theta_{23}$ & 0.44 \ or \ 0.60 \\
    $\phi_{13}$ (rad) & $[0,\,2\pi]$ \\ 
    $\Delta m^2_{21}$ & $7.53\times10^{-5}$~eV$^2$  \\
    $\Delta m^2_{32}$ (normal hierarchy)   & $2.51\times10^{-3}$~eV$^2$  \\
    $\Delta m^2_{32}$ (inverted hierarchy) & $-2.56\times10^{-3}$~eV$^2$ \\ \hline
    $\sin^2\theta_{14}$ & 0.04\\
    $\sin^2\theta_{24}$ & $[0,\,0.006]$\\
    $\sin^2\theta_{34}$ & 0 \\
    $\phi_{14}$ (rad) & equals $\phi_{13}$ when $\sin^2\theta_{23}=0.44$, \ \ \ 
    equals $\phi_{13}+\pi$ when $\sin^2\theta_{23}=0.60$ \\
    $\phi_{34}$ (rad) & unphysical \\
    $\Delta m^2_{41}$ & 3~eV$^2$ \\ \hline
  \end{tabular}
  \label{physpara3}
  \end{center}
\end{table}

\subsection{Experiments}

We assume that J-PARC operates with 1.3~MW beam power, delivering $2.7\times10^{21}$~proton-on-target (POT) flux per year.
We adopt the values in Table~\ref{exppara} as the baseline and detector parameters, where the matter density is approximated to be uniform~\cite{koike,senda}.
We consider three experiments, referred to as `T2HK', `T2HKK' and `T2HK+Oki' in this paper.
The configuration of each experiment is assumed as follows, and is summarized in Table~\ref{exppara}.

\begin{itemize}

\item
In `T2HK', we assume that one 187~kton fiducial volume detector at Kamioka is exposed to
 a neutrino-focusing beam for 2.5~years and to an antineutrino-focusing beam for 2.5~years.
Subsequently, two 187~kton detectors (374~kton in total) at Kamioka are exposed to a neutrino-focusing beam for 5~years and to an antineutrino-focusing beam for 5~years.

\item
In `T2HKK', we assume that one 187~kton detector at Kamioka is exposed to
 a neutrino-focusing beam for 2.5~years and to an antineutrino-focusing beam for 2.5~years.
Subsequently, one 187~kton detector at Kamioka and one 187~kton detector in Korea are exposed to a neutrino-focusing beam for 5~years and to an antineutrino-focusing beam for 5~years.

\item
In `T2HK+Oki', we assume that one 187~kton detector at Kamioka is exposed to
 a neutrino-focusing beam for 2.5~years and to an antineutrino-focusing beam for 2.5~years.
Subsequently, one 187~kton detector at Kamioka and a smaller 100~kton detector at Oki Islands are exposed to a neutrino-focusing beam for 5~years and to an antineutrino-focusing beam for 5~years.

\end{itemize}

\begin{table}[H]
\begin{center}
  \caption{Baseline and detector parameters.
  $L$, FV, $\rho$, `off-axis', `$\nu$-focusing' and `$\bar{\nu}$-focusing' denote the baseline length, the fiducial volume, matter density, off-axis angle, the time of exposure to a neutrino-focusing beam,
  and that to an antineutrino-focusing beam, respectively. }
  \begin{tabular}{|c|c||c|c|c|c|c|c|c|} \hline
    experiment & site & $L$ & FV & $\rho$ & off-axis & $\nu$-focusing & $\bar{\nu}$-focusing \\ \hline
    T2HK & Kamioka & 295~km & 187~kton & 2.60~g/cm$^3$  & 2.5$^{\circ}$ & 2.5~years & 2.5~years \\ 
               & Kamioka & 295~km & 374~kton & 2.60~g/cm$^3$  & 2.5$^{\circ}$ & 5~years & 5~years \\ \hline
    T2HKK & Kamioka & 295~km & 187~kton & 2.60~g/cm$^3$  & 2.5$^{\circ}$ & 2.5~years & 2.5~years \\
             & Kamioka & 295~km & 187~kton & 2.60~g/cm$^3$  & 2.5$^{\circ}$ & 5~years & 5~years \\
          & Korea & 1000~km & 187~kton & 2.90~g/cm$^3$  & 1.0$^{\circ}$ & 5~years & 5~years \\ \hline
    T2HK+Oki & Kamioka & 295~km & 187~kton & 2.60~g/cm$^3$  & 2.5$^{\circ}$ & 2.5~years & 2.5~years \\
    & Kamioka & 295~km & 187~kton & 2.60~g/cm$^3$  & 2.5$^{\circ}$ & 5~years & 5~years \\
          & Oki & 653~km & 100~kton & 2.75~g/cm$^3$  & 1.0$^{\circ}$ & 5~years & 5~years \\ \hline
  \end{tabular}
  \label{exppara}
  \end{center}
\end{table}

\subsection{Number of Events}

The number of $\nu_e+\bar{\nu}_e$ events and the number of $\nu_\mu+\bar{\nu}_\mu$ events
 in the bin of $0.4+(i-1)\cdot 0.05~{\rm GeV}<E<0.4+i \cdot 0.05~{\rm GeV}$
 with a neutrino-focusing beam that are detected at a specific site, denoted by $N_{e,i,{\rm site}}$ and $N_{\mu,i,{\rm site}}$, 
 and those with an antineutrino-focusing beam, denoted by $\widetilde{N}_{e,i,{\rm site}}$ and $\widetilde{N}_{\mu,i,{\rm site}}$,
 are computed as
\begin{align}
{\rm For} \ i=&1,2,3,...,52,
\nn\\
N_{e,i,{\rm site}}&=\int_{0.4+(i-1) \cdot 0.05~{\rm GeV}}^{0.4+i \cdot 0.05~{\rm GeV}} {\rm d}E \
\varepsilon_{e,{\rm site}}\ f_{\sigma e} \left\{ \, f_{\Phi_{\nu}}\Phi_{\nu,{\rm site}}(E) \ P_{\rm site}(\nu_\mu\to\nu_e;E) \ N_{n,{\rm site}} \ 
\sigma(\nu_\ell n \to \ell^- p;E) \right.
\nn\\
&\left. \ \ \ \ \ \ \ \ \ \ \ \ \ \ \ \ \ \ \ \ \ \ \ \ \ \ \ \ \ 
+ f_{\Phi_{\bar{\nu}}}\Phi_{\bar{\nu},{\rm site}}(E)\ P_{\rm site}(\bar{\nu}_\mu \to \bar{\nu}_e;E) \ N_{p,{\rm site}}  \ 
\sigma(\bar{\nu}_\ell p \to \ell^+ n;E) \, \right\},
\\
N_{\mu,i,{\rm site}}&=\int_{0.4+(i-1) \cdot 0.05~{\rm GeV}}^{0.4+i \cdot 0.05~{\rm GeV}} {\rm d}E \
\varepsilon_{\mu,{\rm site}} \ f_{\sigma\mu} \left\{ \, f_{\Phi_{\nu}}\Phi_{\nu,{\rm site}}(E) \ P_{\rm site}(\nu_\mu \to \nu_\mu;E) \ N_{n,{\rm site}} \ \sigma(\nu_\ell n \to \ell^- p;E) \right.
\nn\\
&\left. \ \ \ \ \ \ \ \ \ \ \ \ \ \ \ \ \ \ \ \ \ \ \ \ \ \ \ \ \ 
+ f_{\Phi_{\bar{\nu}}}\Phi_{\bar{\nu},{\rm site}}(E) \ P_{\rm site}(\bar{\nu}_\mu \to \bar{\nu}_\mu;E) \ N_{p,{\rm site}} \ \sigma(\bar{\nu}_\ell p \to \ell^+ n;E) \, \right\},
\\
\widetilde{N}_{e,i,{\rm site}}&={\rm replace} \ (\Phi_{\nu,{\rm site}}(E),\,\Phi_{\bar{\nu},{\rm site}}(E),\,\varepsilon_{e,{\rm site}},\,
f_{\Phi_{\nu}},\,f_{\Phi_{\bar{\nu}}},\,
f_{\sigma e}
) \ {\rm in} \ N_{e,i,{\rm site}} \ 
\nn\\
&\ \ \ \ \ \ \ \ \ \ \ \ \ \ \ \ \ \ \ \ \ \ \ \ \ \ \ \ \ \ \ \ \ \ \ \ \ \ \ {\rm with} \
(\widetilde{\Phi}_{\nu,{\rm site}}(E),\,\widetilde{\Phi}_{\bar{\nu},{\rm site}}(E),\,\widetilde{\varepsilon}_{e,{\rm site}},\,
\widetilde{f}_{\Phi_{\nu}},\,\widetilde{f}_{\Phi_{\bar{\nu}}},\,
\widetilde{f}_{\sigma e}
)
\nn\\
\widetilde{N}_{\mu,i,{\rm site}}&={\rm replace} \ (\Phi_{\nu,{\rm site}}(E),\,\Phi_{\bar{\nu},{\rm site}}(E),\,\varepsilon_{\mu,{\rm site}},\,
f_{\Phi_{\nu}},\,f_{\Phi_{\bar{\nu}}},
f_{\sigma \mu}
) \ {\rm in} \ N_{\mu,i,{\rm site}}
\nn\\
& \ \ \ \ \ \ \ \ \ \ \ \ \ \ \ \ \ \ \ \ \ \ \ \ \ \ \ \ \ \ \ \ \ \ \ \ \ \ {\rm with} \
(\widetilde{\Phi}_{\nu,{\rm site}}(E),\,\widetilde{\Phi}_{\bar{\nu},{\rm site}}(E),\,\widetilde{\varepsilon}_{\mu,{\rm site}},\,
\widetilde{f}_{\Phi_{\nu}},\,\widetilde{f}_{\Phi_{\bar{\nu}}},\,
\widetilde{f}_{\sigma\mu}
)
\nn
\end{align}
 where $\Phi_{\nu,{\rm site}}(E)$, $\Phi_{\bar{\nu},{\rm site}}(E)$ ($\widetilde{\Phi}_{\nu,{\rm site}}(E)$, $\widetilde{\Phi}_{\bar{\nu},{\rm site}}(E)$)
 respectively denote $\nu_\mu$ and $\bar{\nu}_\mu$ fluxes per energy in a neutrino-focusing beam (in an antineutrino-focusing beam) at a specific detector.
$P_{\rm site}$ denotes a transition probability calculated with Eq.~(\ref{prob}) for a specific site.
$N_{n,{\rm site}}$ and $N_{p,{\rm site}}$ are respectively the number of neutrons and protons in the water Cerenkov detector at a specific site.
$\sigma$ denotes the cross sections for the $\nu_\ell n \to \ell^- p$ and $\bar{\nu}_\ell p \to \ell^+ n$ processes.

\begin{figure}[H]
  \begin{center}
    \includegraphics[width=80mm]{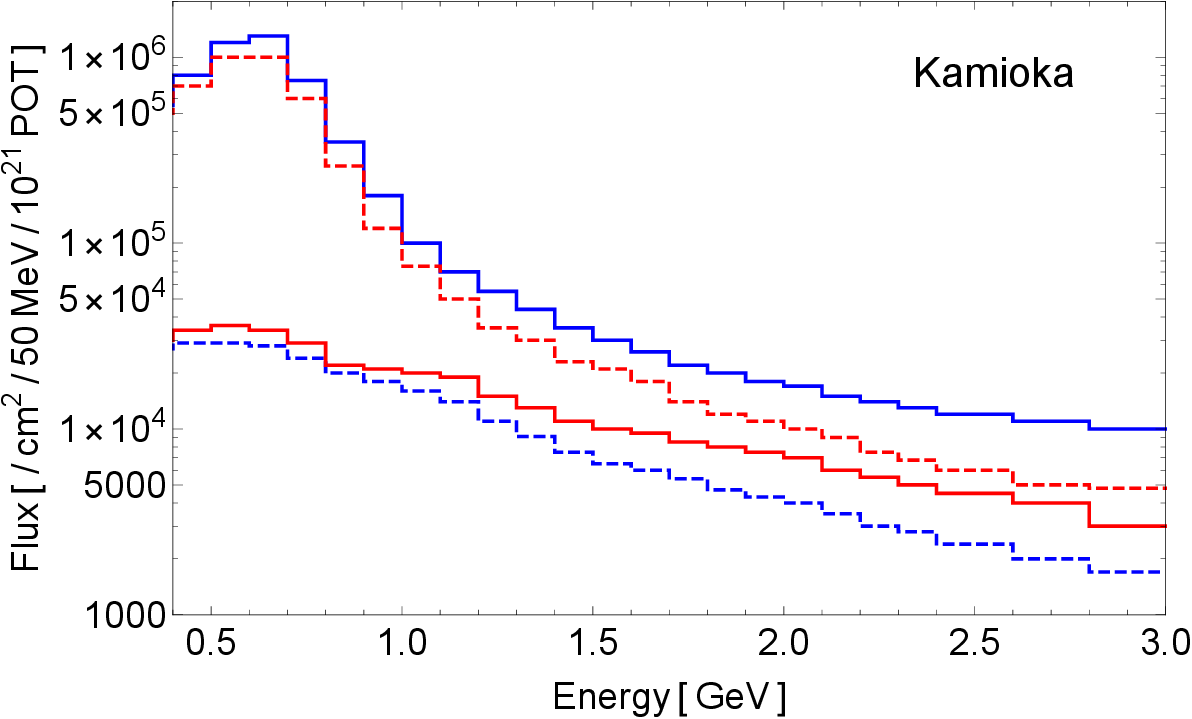}
    \\
    \includegraphics[width=80mm]{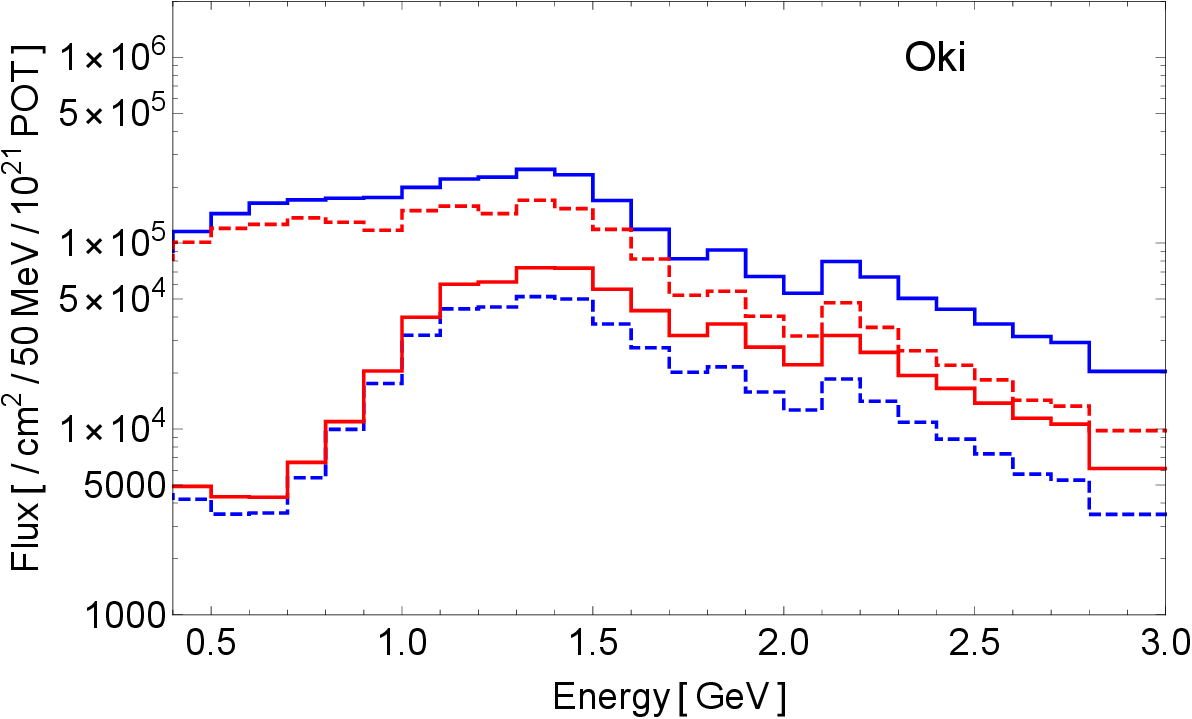}
    \includegraphics[width=80mm]{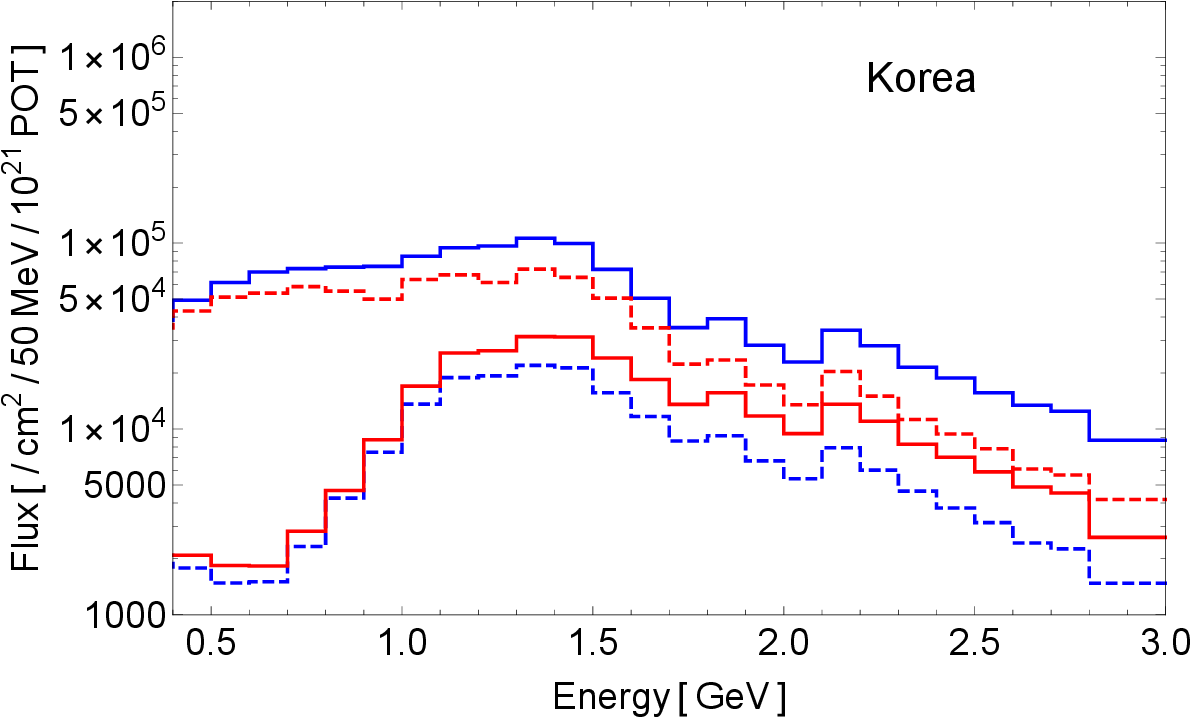}
    \caption{
    Flux of $\nu_\mu$ and $\bar{\nu}_\mu$ in neutrino-focusing and antineutrino-focusing beams detected at
     a water Cerenkov detector at Kamioka, Oki and in Korea if the neutrino oscillations were absent, taken from Ref.~\cite{beam}.
    The baseline length and beam off-axis angle of each detector are as given in Table~\ref{exppara}.
    The upper plot, the lower-left plot and the lower-right plot correspond to Kamioka, Oki and Korea, respectively.
    In the plots, the blue lines correspond to a neutrino-focusing beam and the red lines to an antineutrino-focusing beam.
    The solid lines indicate $\nu_\mu$ flux and the dashed lines $\bar{\nu}_\mu$ flux.
    }
    \label{fluxfig}
  \end{center}
\end{figure}

We employ the result of Ref.~\cite{beam}
 for $\nu_\mu$ and $\bar{\nu}_\mu$ flux
 in neutrino-focusing and antineutrino-focusing beams
 emitted from J-PARC and detected at
 a water Cerenkov detector at Kamioka, Oki and in Korea if the neutrino oscillations were absent.
For reference, we plot the flux in Fig.~\ref{fluxfig}.
The baseline length and beam off-axis angle of each detector are found in Table~\ref{exppara}.
In the plots, the blue lines correspond to a neutrino-focusing beam and the red lines to an antineutrino-focusing beam.
The solid lines indicate $\nu_\mu$ flux and the dashed lines $\bar{\nu}_\mu$ flux.
$\nu_e$ and $\bar{\nu}_e$ components in the beams are neglected in our study.

The cross sections for charged current quasi-elastic scattering between a neutrino and a proton, 
 $\nu_\ell \, n \to \ell^- \, p$, and that between an antineutrino and a neutron, $\bar{\nu}_\ell \, p \to \ell^+ \, n$,
 ($p$ and $n$ denote proton and neutron, respectively, and $\ell$ denotes $e$ or $\mu$)
 are obtained from Ref.~\cite{crosssection}.

$\varepsilon_{e,{\rm site}}$, $\varepsilon_{\mu,{\rm site}}$
 respectively denote the efficiencies for electrons and muons of the far detector at a specific site, in a neutrino-focusing run.
$f_{\Phi_{\nu}}$, $f_{\Phi_{\bar{\nu}}}$ account for the uncertainty of
 neutrino and antineutrino fluxes, respectively, in a neutrino-focusing beam.
$f_{\sigma e}$ ($f_{\sigma \mu}$) accounts for the uncertainty of the weighted sum of charged current quasi-elastic scattering cross sections of $\nu_e$ and $\bar{\nu}_e$ ($\nu_\mu$ and $\bar{\nu}_\mu$) that is estimated from near detector data and theory, in a neutrino-focusing run.
$\widetilde{\varepsilon}_{e,{\rm site}}$, $\widetilde{\varepsilon}_{\mu,{\rm site}}$, 
$\widetilde{f}_{\Phi_{\nu}}$, $\widetilde{f}_{\Phi_{\bar{\nu}}}$,
$\widetilde{f}_{\sigma e}$, $\widetilde{f}_{\sigma \mu}$ are the corresponding quantities in an antineutrino-focusing run.
In this paper, we neglect energy dependence of the above systematic parameters.
For the efficiencies of the far detectors, based on Table~9 of Ref.~\cite{Abe:2015zbg}, we assume
\footnote{
We consider that the same water Cerenkov detector as Kamioka is situated at Oki.
}
\begin{align}
&\varepsilon_{e,{\rm site}}=1\pm0.007, \ \ \ \ \varepsilon_{\mu,{\rm site}}=1\pm0.010, \ \ \ \ 
\widetilde{\varepsilon}_{e,{\rm site}}=1\pm0.017, \ \ \ \ \widetilde{\varepsilon}_{\mu,{\rm site}}=1\pm0.011.
\label{efficiency}\\
&\ \ \ \ \ \ \ \ \ \ \ \ \ \ \ \ \ \ \ \ \ \varepsilon_{e,{\rm site}} \ {\rm and} \ \widetilde{\varepsilon}_{e,{\rm site}} \ {\rm at \ each \ site \ are \ maximally \ correlated.}
\nn\\
&\ \ \ \ \ \ \ \ \ \ \ \ \ \ \ \ \ \ \ \ \ \varepsilon_{\mu,{\rm site}} \ {\rm and} \ \widetilde{\varepsilon}_{\mu,{\rm site}} \ {\rm at \ each \ site \ are \ maximally \ correlated.}
\nn
\end{align}
For the flux uncertainty, based on Fig.~9 of Ref.~\cite{Abe:2015zbg}, we approximate
\begin{align}
f_{\Phi_{\nu}}&=1\pm0.1, \ \ \ f_{\Phi_{\bar{\nu}}}=1\pm0.08, \ \ \ \widetilde{f}_{\Phi_{\nu}}=1\pm0.1, \ \ \ \widetilde{f}_{\Phi_{\bar{\nu}}}=1\pm0.1.
\label{fphi}                         
\end{align}
For the cross section uncertainty, based on Table~9 of Ref.~\cite{Abe:2015zbg}, we assume 
\begin{align}
f_{\sigma e}&=1\pm\sqrt{0.030^2+0.012^2}, \ \ \ f_{\sigma \mu}=1\pm\sqrt{0.028^2+0.015^2},
\nn\\
\widetilde{f}_{\sigma e}&=1\pm\sqrt{0.056^2+0.020^2}, \ \ \ \widetilde{f}_{\sigma \mu}=1\pm\sqrt{0.042^2+0.014^2}.
\label{fsigma}                         
\end{align}
All uncertainties above are assumed to be uncorrelated unless otherwise stated.
\\

\subsection{$\chi^2$ analysis}

We perform a $\chi^2$ fit of the numbers of events in the bins with neutrino-focusing and antineutrino-focusing beams measured at Kamioka/Korea/Oki, $N_{e,i,{\rm site}}$, $N_{\mu,i,{\rm site}}$, $\widetilde{N}_{e,i,{\rm site}}$
and $\widetilde{N}_{\mu,i,{\rm site}}$ $(i=1,2,...,52;$ site=Kamioka, Korea, Oki).
In the fitting analysis, we vary $\sin^2\theta_{23}$, $\phi_{13}$, $\phi_{14}$ and $\phi_{34}$ freely, 
 while using true values for the other standard mixing angles $\theta_{12},\theta_{13}$, 
 the sterile-active mixing angles $\theta_{14},\theta_{24},\theta_{34}$, 
 and the mass differences $\Delta m^2_{21},|\Delta m^2_{32}|,\Delta m^2_{41}$.
Such an analysis is justifiable because 
 $\theta_{12},\theta_{13},\theta_{14}$, $\theta_{24},\theta_{34}$, $\Delta m^2_{21},|\Delta m^2_{32}|,\Delta m^2_{41}$
 can be measured to a good accuracy in short baseline experiments and atmospheric neutrino measurements,
 while the octant of $\theta_{23}$ and the CP phases $\phi_{13},\phi_{14},\phi_{34}$
 can be measured precisely only in long baseline experiments.
Additionally, we fit the normal hierarchy events by assuming the normal hierarchy, and 
 the inverted hierarchy events by assuming the inverted hierarchy,
 because the mass hierarchy may have been determined before T2HK starts operation.

We take into account the uncertainties of the efficiencies of far detectors
 $\varepsilon_{e,{\rm site}},\varepsilon_{\mu,{\rm site}},\widetilde{\varepsilon}_{e,{\rm site}},\widetilde{\varepsilon}_{\mu,{\rm site}}$, 
 those of neutrino-focusing and antineutrino-focusing beam fluxes
 $f_{\Phi_{\nu}},f_{\Phi_{\bar{\nu}}},f_{\widetilde{\Phi}_{\nu}},f_{\widetilde{\Phi}_{\bar{\nu}}}$, 
 and those of the cross sections (estimated from near detector data and theory)
 $f_{\sigma e},f_{\sigma \mu},\widetilde{f}_{\sigma e},\widetilde{f}_{\sigma \mu}$, which constitute systematic uncertainties.
We vary the above parameters, with the inclusion of the maximal correlation of $\varepsilon_{e,{\rm site}}$ and $\widetilde{\varepsilon}_{e,{\rm site}}$,
 and that of $\varepsilon_{\mu,{\rm site}}$ and $\widetilde{\varepsilon}_{\mu,{\rm site}}$.
Eventually, the $\chi^2$ function is given by
\begin{align}
&\chi^2(\sin^2\theta_{23},\,\phi_{13},\,\phi_{14},\,\phi_{34},\,\Pi_{\rm sys})=\sum_{{\rm site}}\sum_{i=1}^{52}
\nn\\
&\left[\frac{\left(N_{e,i,{\rm site}} - N_{e,i,{\rm site}}^{\rm pred}(\sin^2\theta_{23},\phi_{13},\phi_{14},\phi_{34},\Pi_{\rm sys})\right)^2}{N_{e,i,{\rm site}}}
+\frac{\left( N_{\mu,i,{\rm site}} - N_{\mu,i,{\rm site}}^{\rm pred}(\sin^2\theta_{23},\phi_{13},\phi_{14},\phi_{34},\Pi_{\rm sys})\right)^2}{N_{\mu,i,{\rm site}}}\right.
\nn\\
&\left.\ \ +
\frac{\left( \widetilde{N}_{e,i,{\rm site}} - \widetilde{N}_{e,i,{\rm site}}^{\rm pred}(\sin^2\theta_{23},\phi_{13},\phi_{14},\phi_{34},\Pi_{\rm sys}) \right)^2}{\widetilde{N}_{e,i,{\rm site}}}
+\frac{\left( \widetilde{N}_{\mu,i,{\rm site}} - \widetilde{N}_{\mu,i,{\rm site}}^{\rm pred}(\sin^2\theta_{23},\phi_{13},\phi_{14},\phi_{34},\Pi_{\rm sys}) \right)^2}{\widetilde{N}_{\mu,i,{\rm site}}} \, \right]
\nn\\
&+\sum_{{\rm site}}
\left\{
\left(\frac{\varepsilon_{e,{\rm site}}-1}{0.007}\right)^2+\left(\frac{\varepsilon_{\mu,{\rm site}}-1}{0.010}\right)^2\right\}
\nn\\
&+\left( \frac{f_{\Phi_\nu}-1}{0.1} \right)^2+\left( \frac{f_{\Phi_{\bar{\nu}}}-1}{0.08} \right)^2
+\left(\frac{\widetilde{f}_{\Phi_\nu}-1}{0.1}\right)^2+\left(\frac{\widetilde{f}_{\Phi_{\bar{\nu}}}-1}{0.1}\right)^2
\nn\\
&+\left(\frac{f_{\sigma e}-1}{0.032}\right)^2+\left(\frac{f_{\sigma \mu}-1}{0.032}\right)^2
+\left(\frac{\widetilde{f}_{\sigma e}-1}{0.059}\right)^2+\left(\frac{\widetilde{f}_{\sigma \mu}-1}{0.044}\right)^2
\label{chi2}
\\
&\ \ \ \ \ \ \ \ \ \  \ \ \ \ \ \ \ \ \ \ \ \ {\rm where} \ \ \Pi_{\rm sys}=\left\{\varepsilon_{e,{\rm site}},\varepsilon_{\mu,{\rm site}},f_{\Phi_{\nu}},f_{\Phi_{\bar{\nu}}},\widetilde{f}_{\Phi_{\nu}},\widetilde{f}_{\Phi_{\bar{\nu}}},f_{\sigma e},f_{\sigma \mu},\widetilde{f}_{\sigma e},\widetilde{f}_{\sigma \mu}\right\}
\nn\\
&\ \ \ \ \ \ \ \ \ \  \ \ \ \ \ \ \ \ \ \  \ \ \ \ {\rm and} \ \ 
\widetilde{\varepsilon}_{e,{\rm site}}-1=\frac{0.017}{0.007}(\varepsilon_{e,{\rm site}}-1), \ \ \ \ \ 
\widetilde{\varepsilon}_{\mu,{\rm site}}-1=\frac{0.011}{0.010}(\varepsilon_{\mu,{\rm site}}-1).
\label{chi2app}
\end{align} 
Here $N_{e,i,{\rm site}}$, $N_{\mu,i,{\rm site}}$, $\widetilde{N}_{e,i,{\rm site}}$
and $\widetilde{N}_{\mu,i,{\rm site}}$ are calculated for the central values of Eqs.~(\ref{efficiency}),(\ref{fphi}),(\ref{fsigma}).
$N^{\rm pred}(\sin^2\theta_{23},\phi_{13},\phi_{14},\phi_{34},\Pi_{\rm sys})$ denotes the number of events calculated as a function of $\sin^2\theta_{23}$, $\phi_{13}$, $\phi_{14}$, $\phi_{34}$ and the systematic parameters $\Pi_{\rm sys}=\left\{\varepsilon_{e,{\rm site}},\varepsilon_{\mu,{\rm site}},f_{\Phi_{\nu}},f_{\Phi_{\bar{\nu}}},\widetilde{f}_{\Phi_{\nu}},\widetilde{f}_{\Phi_{\bar{\nu}}},f_{\sigma e},f_{\sigma \mu},\widetilde{f}_{\sigma e},\widetilde{f}_{\sigma \mu}\right\}$,
 using the true values of $\theta_{12},\theta_{13},\Delta m^2_{21},\vert\Delta m^2_{32}\vert,\theta_{14},\theta_{24},\theta_{34}$ and the true mass hierarchy.
The values of $\widetilde{\varepsilon}_{e,{\rm site}},\widetilde{\varepsilon}_{\mu,{\rm site}}$
 are calculated from $\varepsilon_{e,{\rm site}},\varepsilon_{\mu,{\rm site}}$ through Eq.~(\ref{chi2app}),
 since they are maximally correlated.

In this paper, we are interested in the danger of wrong measurement of the $\theta_{23}$ octant.
Thus, we evaluate the significance of rejecting the wrong octant, $\sqrt{\Delta \chi^2}$, which is the square root of the difference between the minimum of $\chi^2$ for $\sin^2\theta_{23}>1/2$ and that for $\sin^2\theta_{23}<1/2$.
Specifically, for each benchmark with $\sin^2\theta_{23}=$0.44 or 0.60, $\Delta \chi^2$ is given by
\begin{align}
&\Delta \chi^2=\min_{\sin^2\theta_{23}>1/2}
\chi^2(\sin^2\theta_{23},\phi_{13},\phi_{14},\phi_{34},\Pi_{\rm sys})-\min_{\sin^2\theta_{23}<1/2}\chi^2(\sin^2\theta_{23},\phi_{13},\phi_{14},\phi_{34},\Pi_{\rm sys})
\nn\\
&\ \ \ \ \ \ \ \ \ \ \ \ \  \ \ \ \ \ \ \ \ \ \  \ \ \ \ \ \ \ \ \ \ \ \ \ \ \ \ \ \ \ \  \ \ \ \ \ \ \ \ \ \  \ \ \ \ \ \ \ \ \ \ {\rm when} \ {\rm true} \ \sin^2\theta_{23} \ {\rm is} \ 0.44,
\label{chisq044}\\
&\Delta \chi^2=\min_{\sin^2\theta_{23}<1/2}
\chi^2(\sin^2\theta_{23},\phi_{13},\phi_{14},\phi_{34},\Pi_{\rm sys})-\min_{\sin^2\theta_{23}>1/2}\chi^2(\sin^2\theta_{23},\phi_{13},\phi_{14},\phi_{34},\Pi_{\rm sys})
\nn\\
&\ \ \ \ \ \ \ \ \ \  \ \ \ \ \ \ \ \ \ \  \ \ \ \ \ \ \ \ \ \ \ \ \ \ \ \ \ \ \ \  \ \ \ \ \ \ \ \ \ \  \ \ \ \ \ \ \ \ \ \ \ \ \ {\rm when} \ {\rm true} \ \sin^2\theta_{23} \ {\rm is} \ 0.60.
\label{chisq060}
\end{align}
We have confirmed that for both $\sin^2\theta_{23}=$0.44 and 0.60,
 the $\chi^2$ function has local minima in both $\sin^2\theta_{23}>1/2$ and $\sin^2\theta_{23}<1/2$ regions.
\\

\subsection{Results}

Before going to the main results, we show in Table~\ref{numbers} the total numbers of $\nu_\mu+\bar{\nu}_\mu$ events at the Kamioka, Oki and Korea detectors in T2HK, T2HKK and T2HK+Oki experiments when there were {\it no oscillations},
corresponding to the configurations of Table~\ref{exppara} and the assumption that J-PARC operates with 1.3~MW beam power.
\begin{table}[H]
\begin{center}
  \caption{The total number of $\nu_\mu+\bar{\nu}_\mu$ events at each detector in each experiment when there were {\it no oscillations}, 
  corresponding to the configurations of Table~\ref{exppara} and J-PARC operation with 1.3~MW beam power.
  }
  \begin{tabular}{|c|c||c|c|} \hline
      experiment & detector &  event number with  & event number with \\ 
                        &                  &  neutrino-focusing beam ($\times10^4$) & antineutrino-focusing beam  ($\times10^4$) \\ \hline
     T2HK          & Kamioka  & 13  & 4.7 \\ \hline
     T2HKK        & Kamioka  & 7.9  & 2.8 \\
                         & Korea      &  1.6 &  0.86 \\ \hline
     T2HK+Oki  & Kamioka  & 7.9 & 2.8 \\
                         & Oki           &  2.0 &  1.1 \\ \hline
  \end{tabular}
  \label{numbers}
  \end{center}
\end{table}

\begin{figure}[H]
  \begin{center}
    \includegraphics[width=55mm]{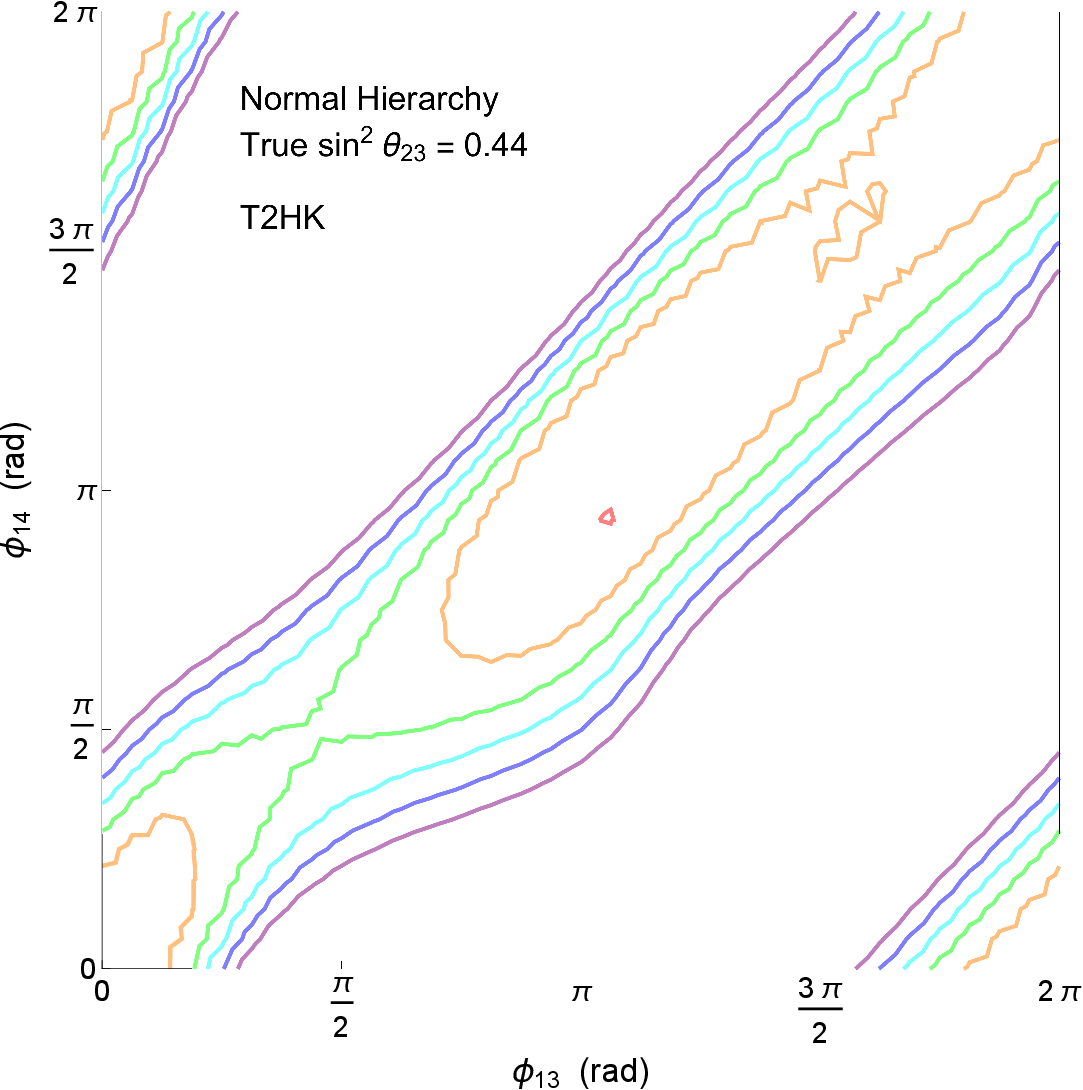}
    \includegraphics[width=55mm]{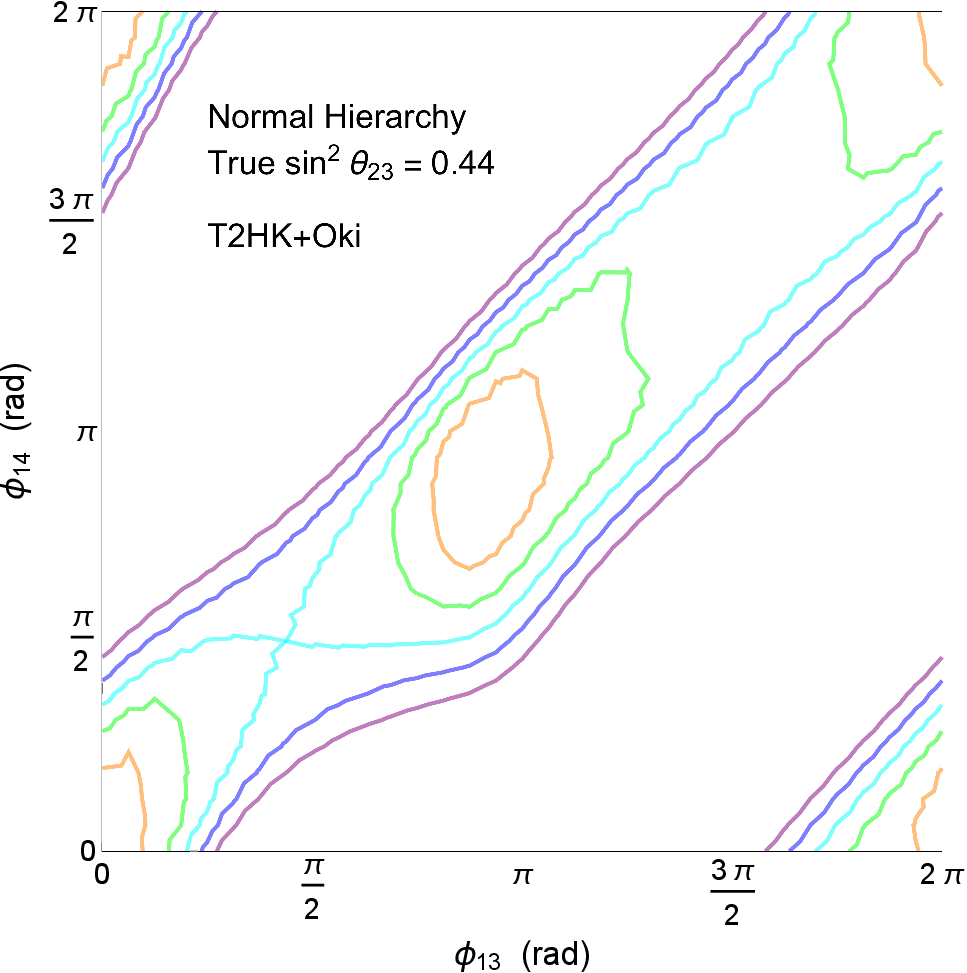}
    \includegraphics[width=55mm]{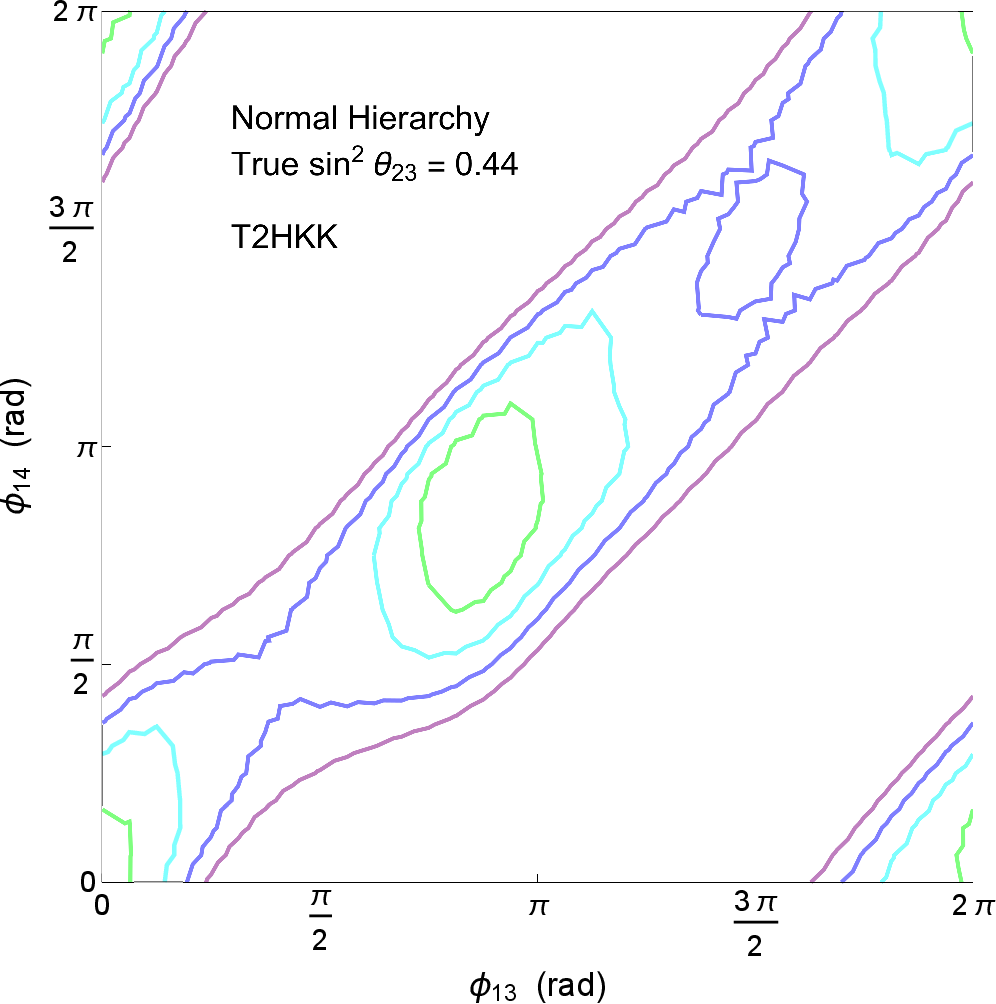}
    \\
        \includegraphics[width=55mm]{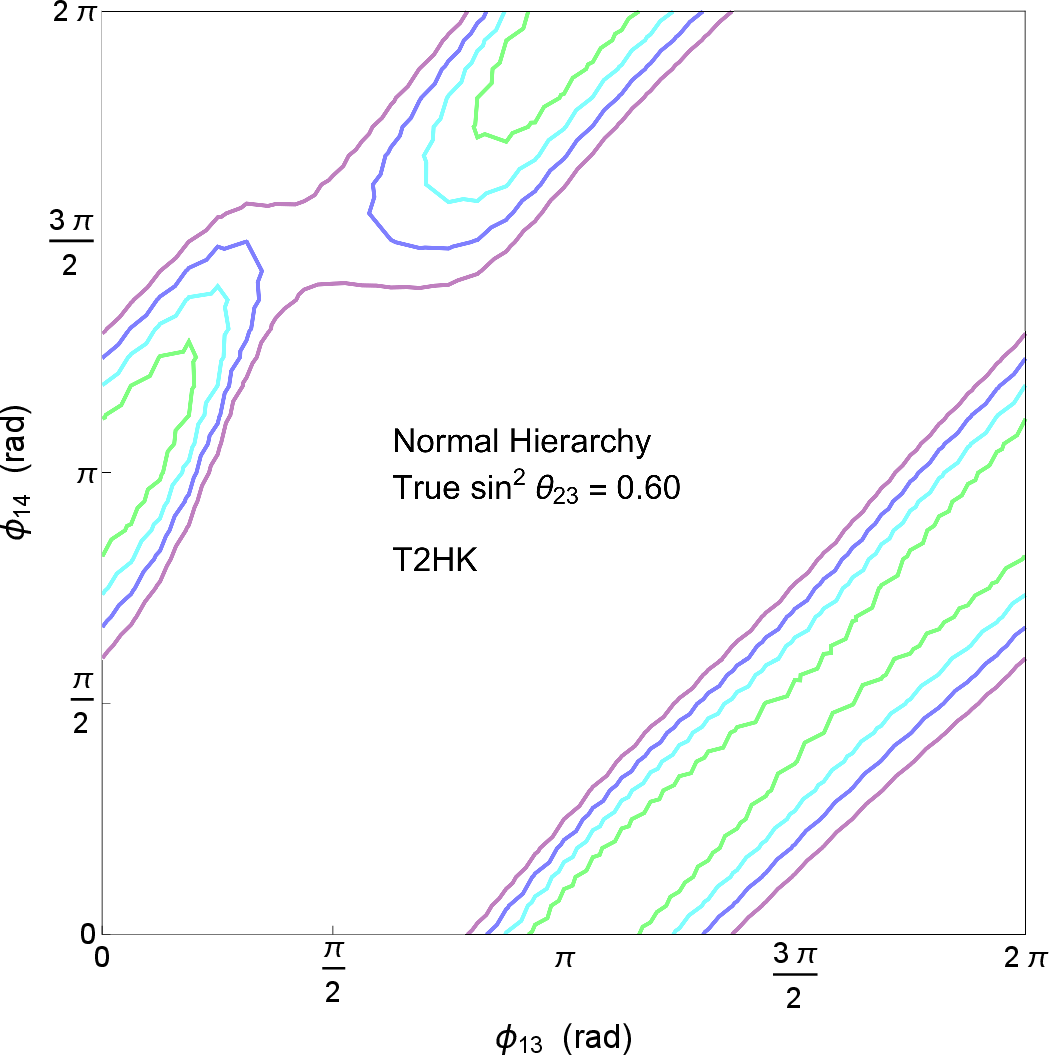}
    \includegraphics[width=55mm]{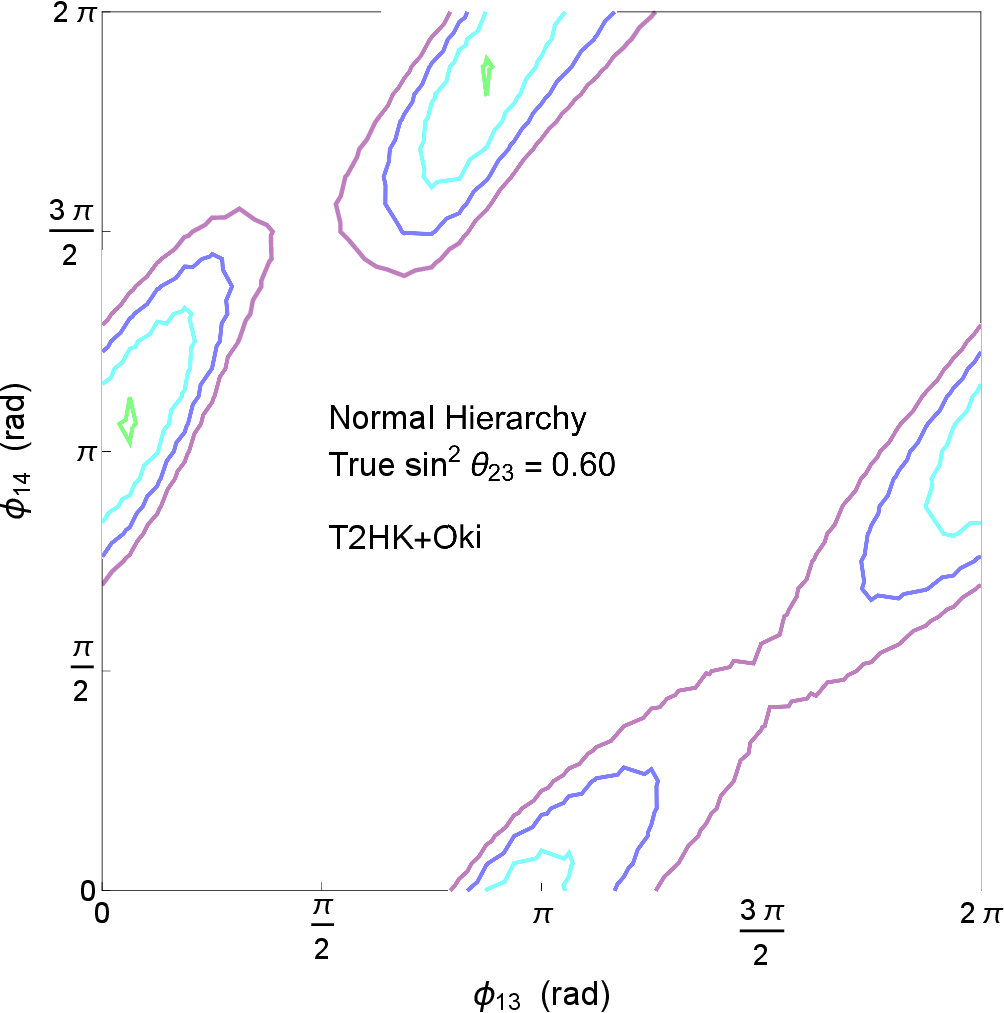}
    \includegraphics[width=55mm]{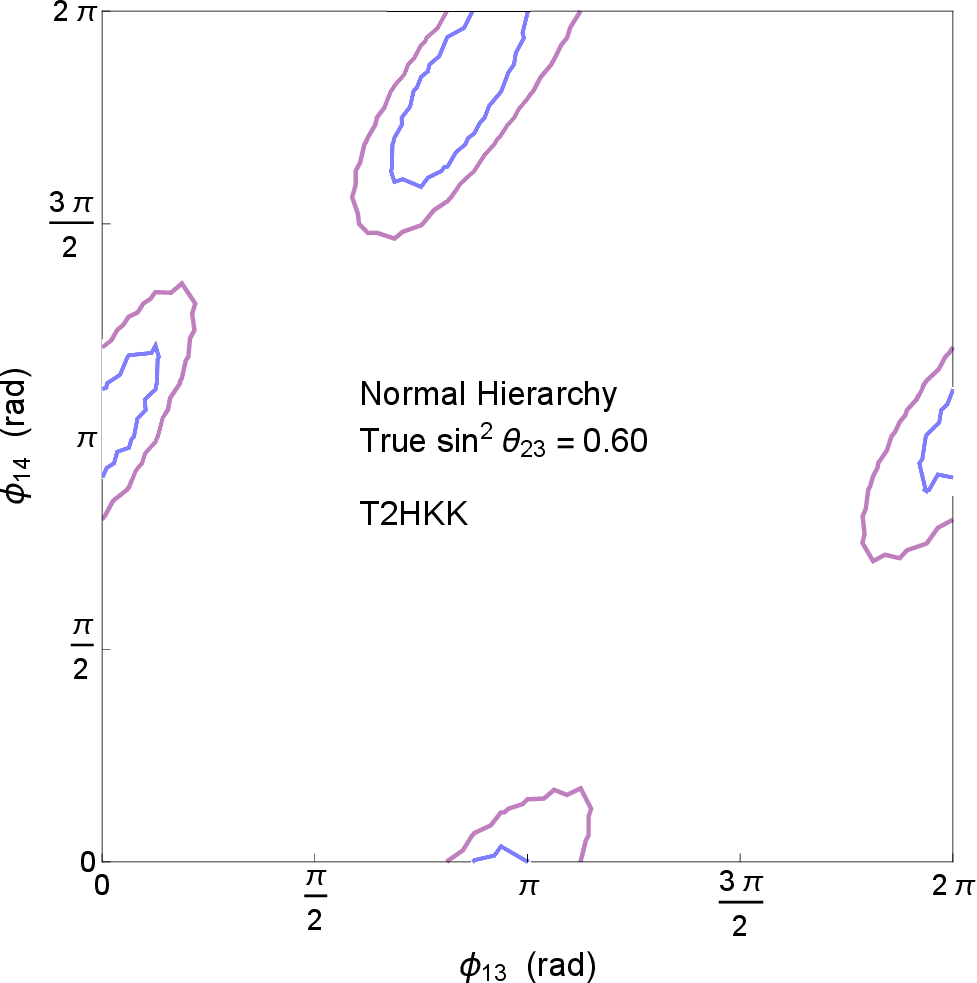}
    \caption{
    Significance of rejecting the wrong $\theta_{23}$ octant when the true $\sin^2\theta_{23}$ is 0.44,
     Eq.~(\ref{chisq044}) (upper three panels), and that when the true $\sin^2\theta_{23}$ is 0.60,
     Eq.~(\ref{chisq060}) (lower three panels),
     in T2HK, T2HKK and T2HK+Oki experiments, in the first benchmark Table~\ref{physpara}.
     The mass hierarchy is normal.
    The red, orange, green, cyan, blue and purple contours correspond to
 $\Delta \chi^2=1^2,\ 1.5^2,\ 2^2,\ 2.5^2,\ 3^2,\ 3.5^2$, respectively.
    }
    \label{nh}
  \end{center}
\end{figure}

\begin{figure}[H]
  \begin{center}
    \includegraphics[width=55mm]{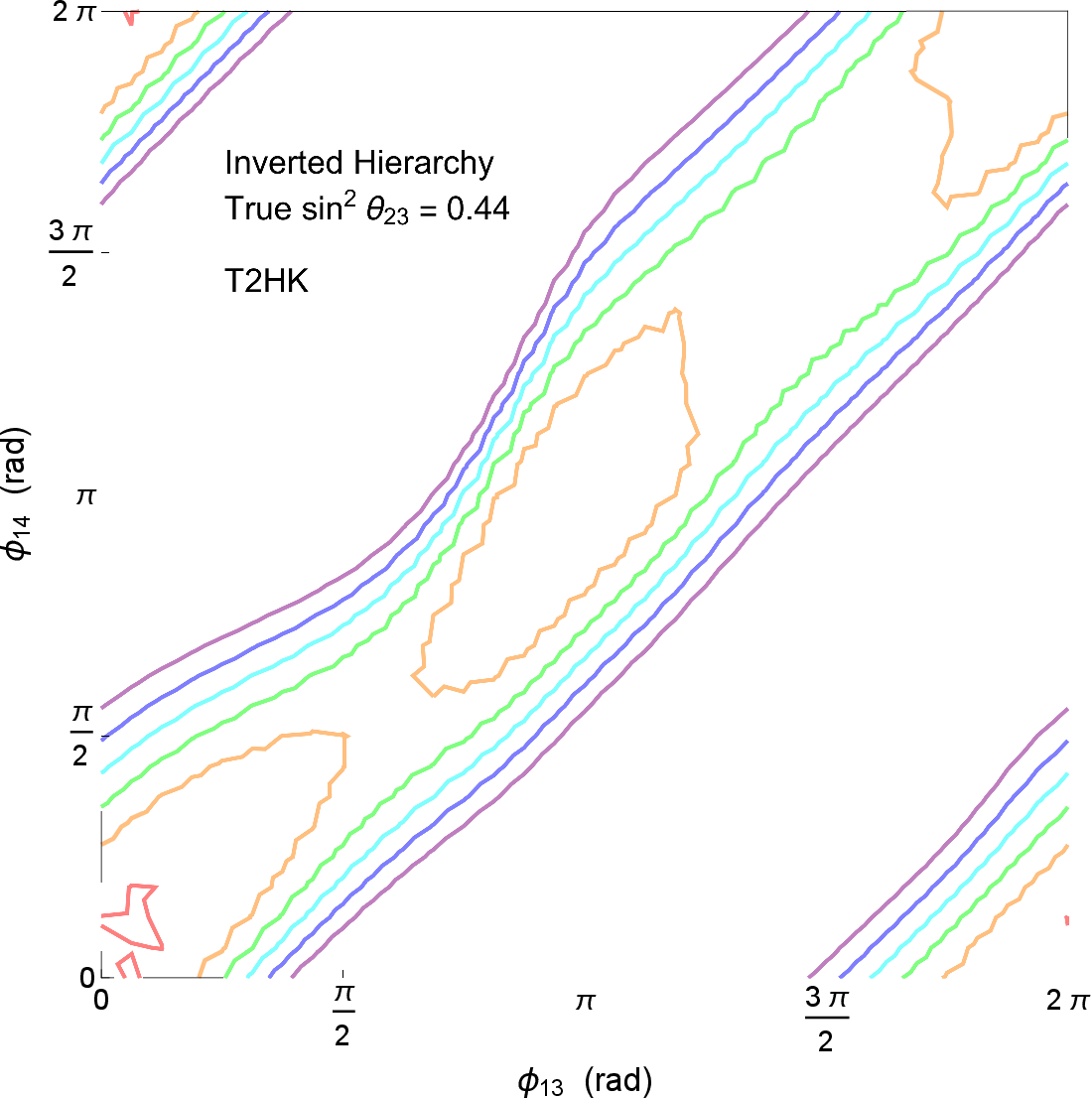}
    \includegraphics[width=55mm]{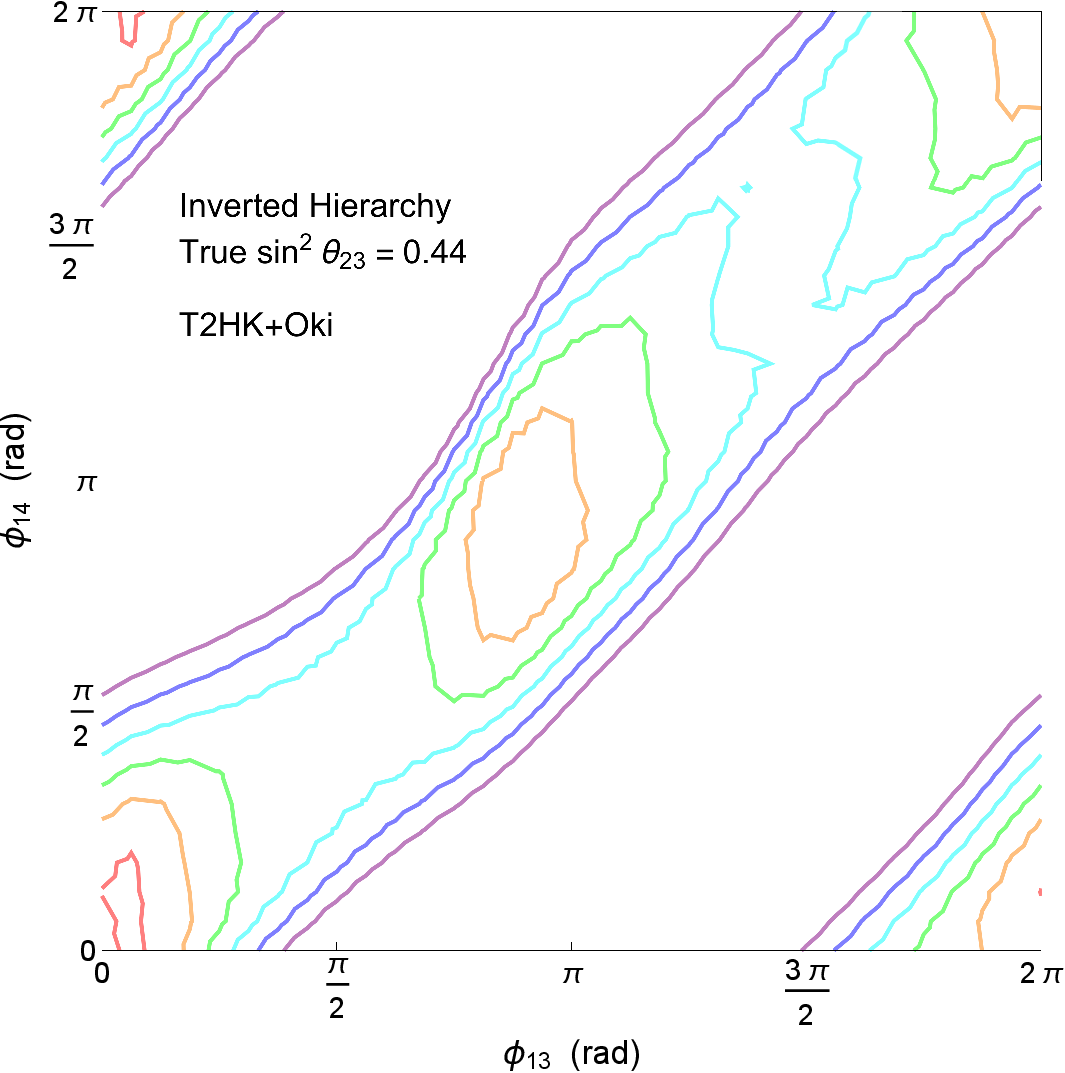}
    \includegraphics[width=55mm]{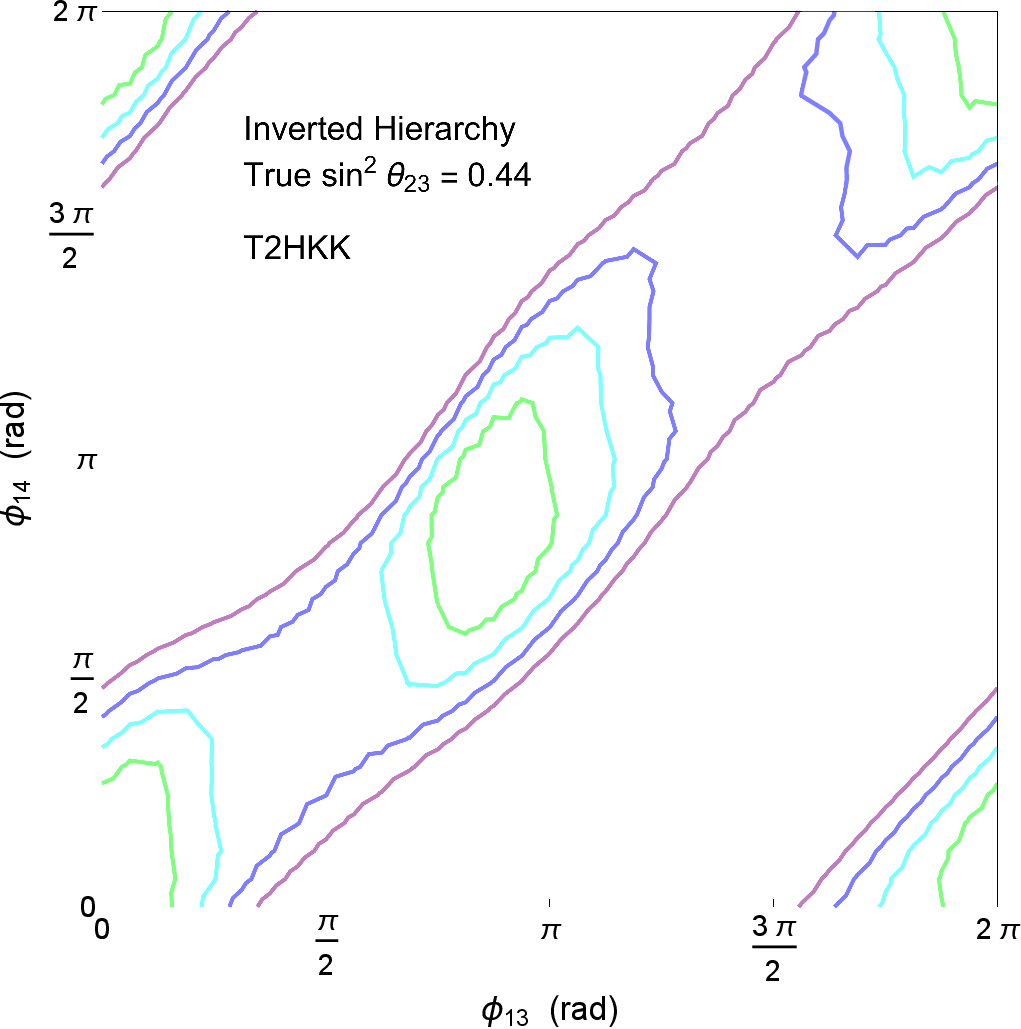}
    \\
        \includegraphics[width=55mm]{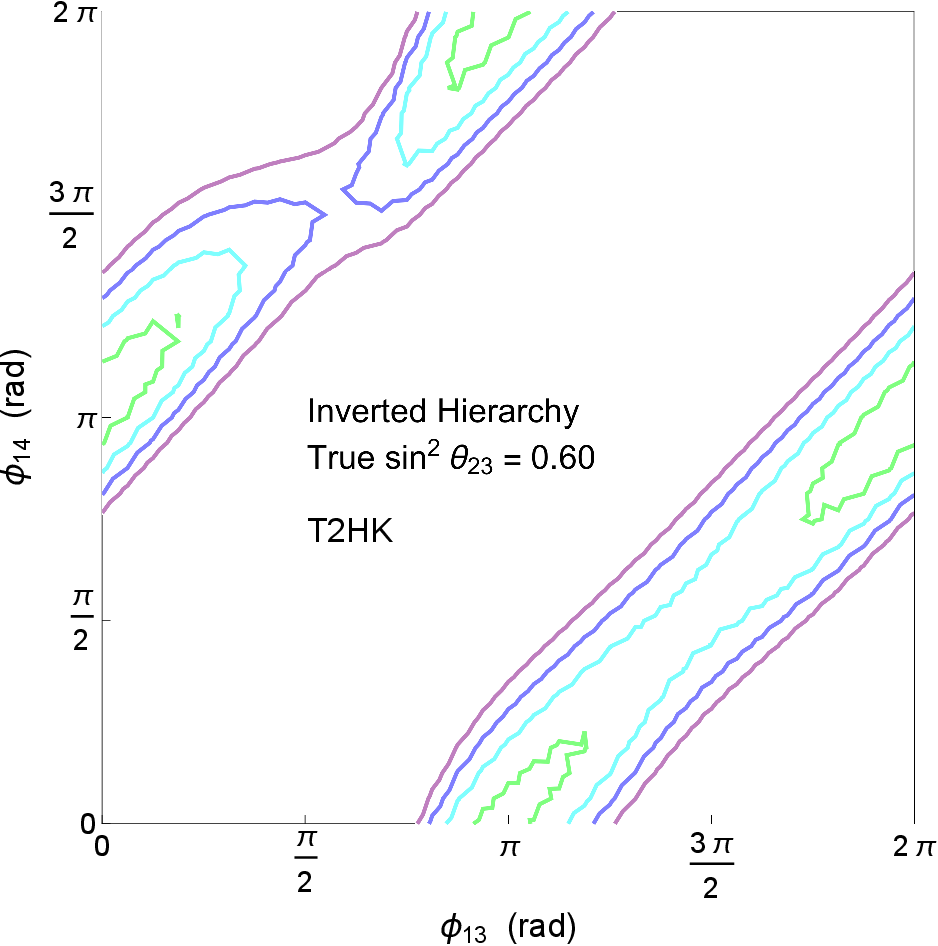}
    \includegraphics[width=55mm]{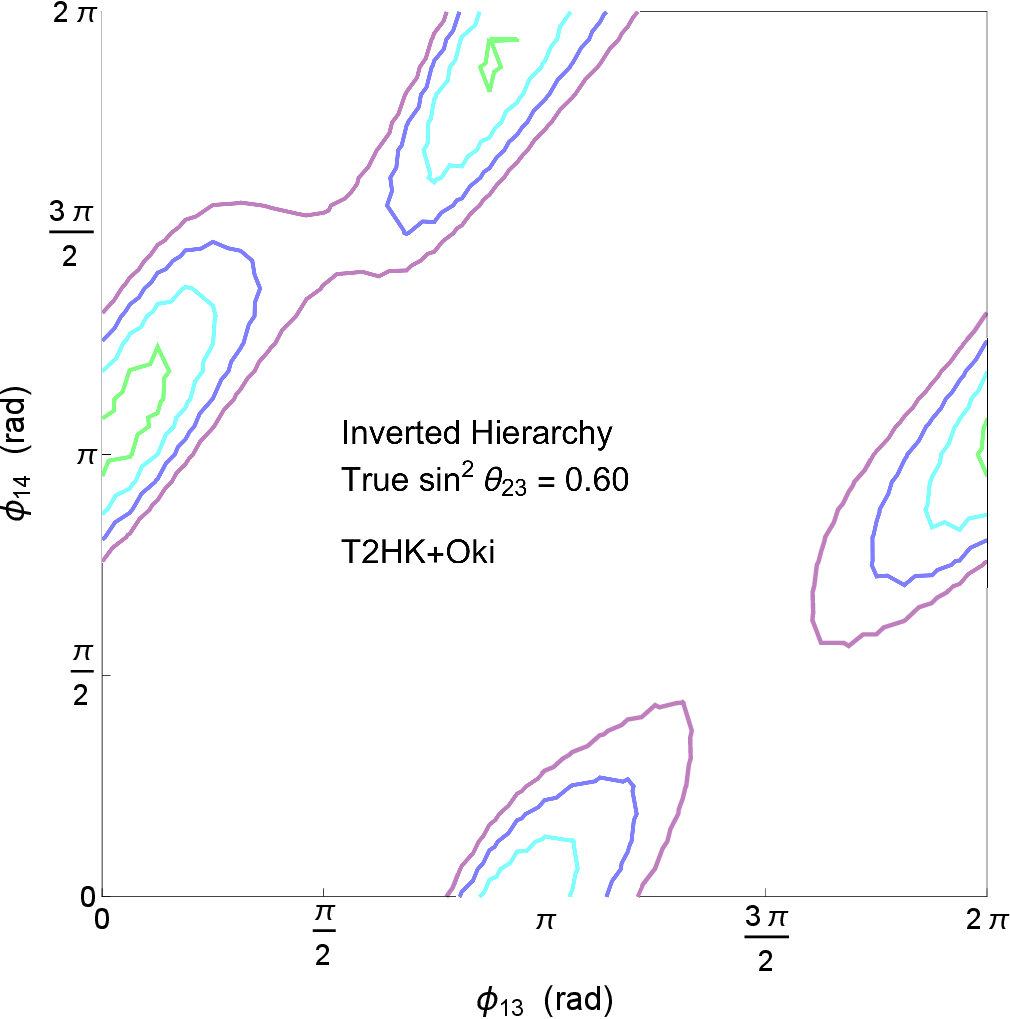}
    \includegraphics[width=55mm]{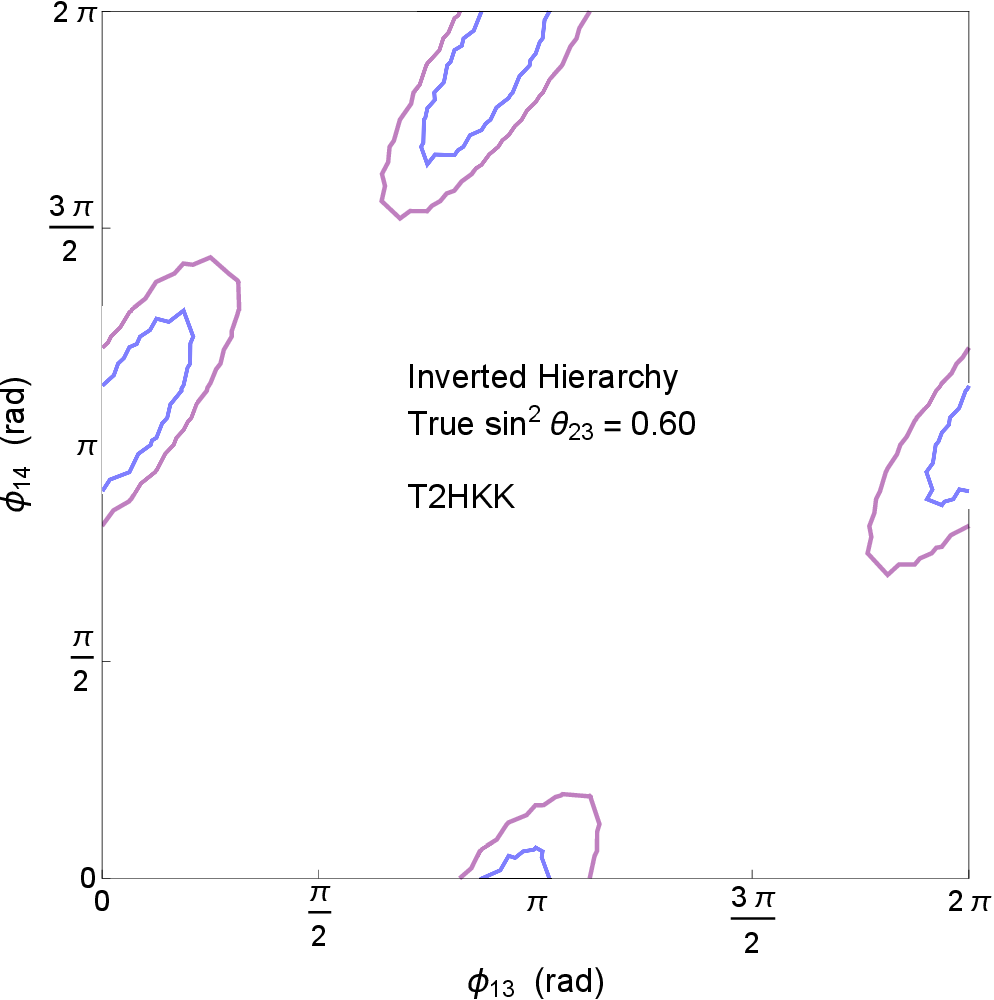}
    \caption{
    Same as Fig.~\ref{nh} except that the mass hierarchy is inverted.
    }
    \label{ih}
  \end{center}
\end{figure}

Now we present $\Delta \chi^2$ with $\sin^2\theta_{23}=$0.44 Eq.~(\ref{chisq044})
 and that with $\sin^2\theta_{23}=$0.60 Eq.~(\ref{chisq060}) in the first benchmark Table~\ref{physpara}.
The results are presented as a function of $(\phi_{13},\phi_{14})$ and shown separately for the normal and inverted mass hierarchies.
Fig.~\ref{nh} is the result when the mass hierarchy is normal,
 and Fig.~\ref{ih} is the result when the mass hierarchy is inverted.
The red, orange, green, cyan, blue and purple contours correspond to
 $\Delta \chi^2=1^2,\ 1.5^2,\ 2^2,\ 2.5^2,\ 3^2,\ 3.5^2$, respectively.

The following observations are made from Figs.~\ref{nh},\ref{ih}.

\begin{itemize}

\item
In T2HK, the significance of rejecting the wrong $\theta_{23}$ octant can decrease as low as $\sqrt{\Delta \chi^2}<1.5$
 when $\sin^2\theta_{23}=0.44$, and as low as $\sqrt{\Delta \chi^2}<2$ when $\sin^2\theta_{23}=0.60$,
 in a wide region of the $(\phi_{13},\,\phi_{14})$ space for both mass hierarchies.
This corroborates the claim of Ref.~\cite{Agarwalla:2016xlg}.

\item
In T2HK, when $\sin^2\theta_{23}=0.44$, 
 the decrease of the significance is most prominent for $\phi_{13}\simeq\phi_{14}$, and in particular for  
 $(\phi_{13},\,\phi_{14})=(0,0),~(\pi,\pi)$.
When $\sin^2\theta_{23}=0.60$, 
 the decrease of the significance is most prominent for $\phi_{13}\simeq\phi_{14}+\pi$, and in particular for  
 $(\phi_{13},\,\phi_{14})=(0,\pi),~(\pi,0)$.
The above results are quite consistent with Eqs.~(\ref{main}),(\ref{sinsq})
\footnote{
The term Eq.~(\ref{sin2}) is small around the energy peak at the Kamioka detector and so plays no role in the following argument.
}.
Since $B\geq0$ in our phase definition, the second part of Eq.~(\ref{sinsq}) is positively maximal when $\phi_{13}\simeq\phi_{14}$,
 and is negatively maximal when $\phi_{13}\simeq\phi_{14}+\pi$.
Hence, this part is most likely to lead to
 wrong $\theta_{23}$ octant measurement when $\sin^2\theta_{23}<1/2$ and $\phi_{13}\simeq\phi_{14}$,
 and when $\sin^2\theta_{23}>1/2$ and $\phi_{13}\simeq\phi_{14}+\pi$.
Moreover, when $\phi_{13}=0,\,\pi$, the first part of Eq.~(\ref{sinsq}) vanishes and thus the combination of neutrino and antineutrino oscillations cannot resolve the above degeneracy.

\item
T2HKK and T2HK+Oki give a larger significance of rejecting the wrong $\theta_{23}$ octant than T2HK 
 in all cases, 
 in spite of their smaller statistics than T2HK (see Table~\ref{numbers}).
This confirms the qualitative argument of Section~2 of the present paper.

\item
Comparing T2HKK and T2HK+Oki, we observe that T2HKK gives a larger significance, in spite of the fact that T2HKK has smaller statistics than T2HK+Oki (see Table~\ref{numbers}).
The reason that T2HK+Oki, in spite of its larger statistics, shows a smaller improvement of the octant sensitivity than T2HKK
 is understood as follows:
As shown in Fig.~\ref{fluxfig}, under our assumption that the beam off-axis angle at Oki is 1.0$^\circ$,
 the flux at Oki has a (broad) peak around $E\simeq1.4$~GeV.
Thus, the value of $L/E$ at the flux peak is 653~km$/$1.4~GeV for Oki, and it is 295~km$/$0.6~GeV for Kamioka.
Since 653~km$/$1.4~GeV and 295~km$/$0.6~GeV are close values,
  a major portion of (anti)neutrinos oscillate in a similar manner at Oki and Kamioka,
  which spoils our attempt to resolve the degeneracy between Eq.~(\ref{main}) and Eqs.~(\ref{sinsq}),(\ref{sin2})
  by combining different baselines.

\end{itemize}

\begin{figure}[H]
  \begin{center}
    \includegraphics[width=55mm]{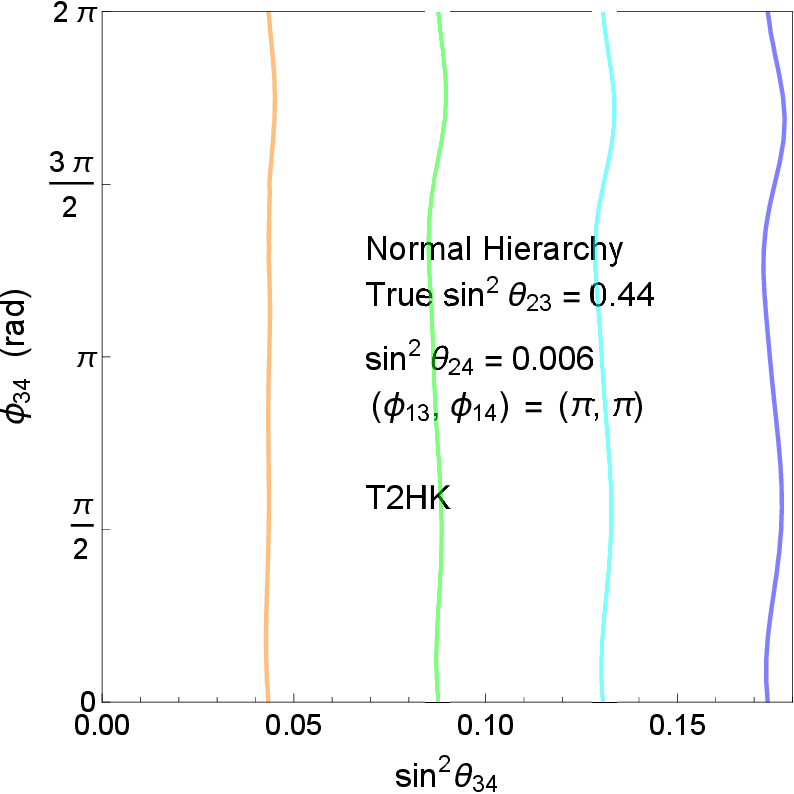}
    \includegraphics[width=55mm]{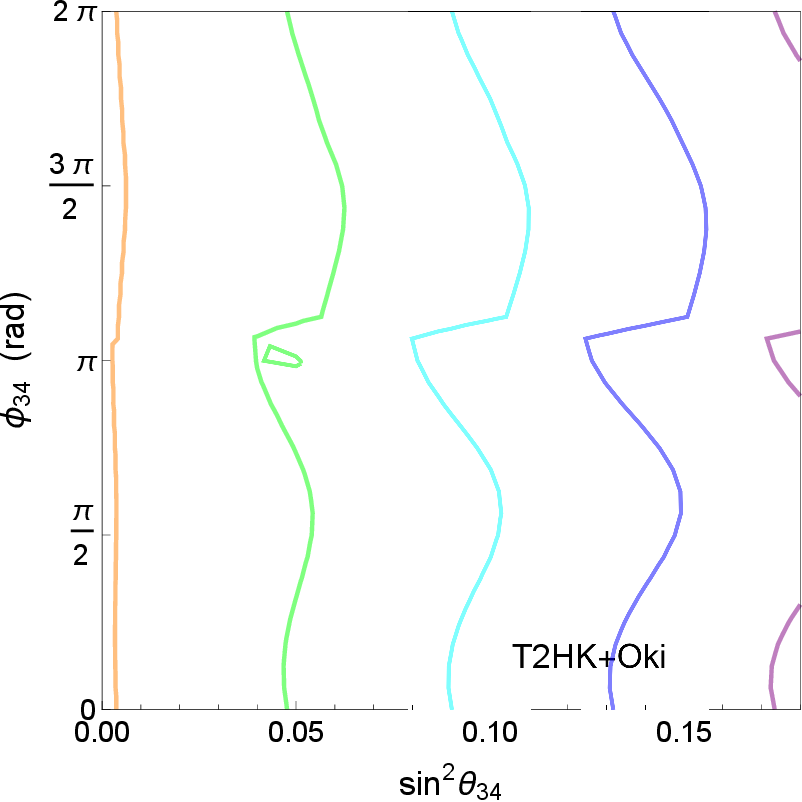}
    \includegraphics[width=55mm]{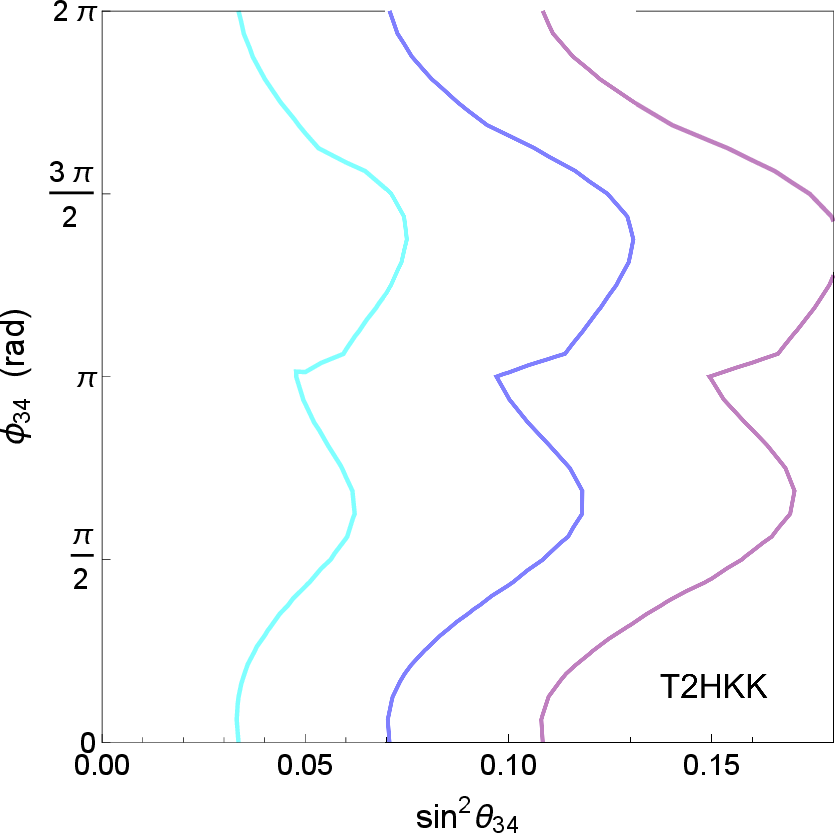}
    \\
     \includegraphics[width=55mm]{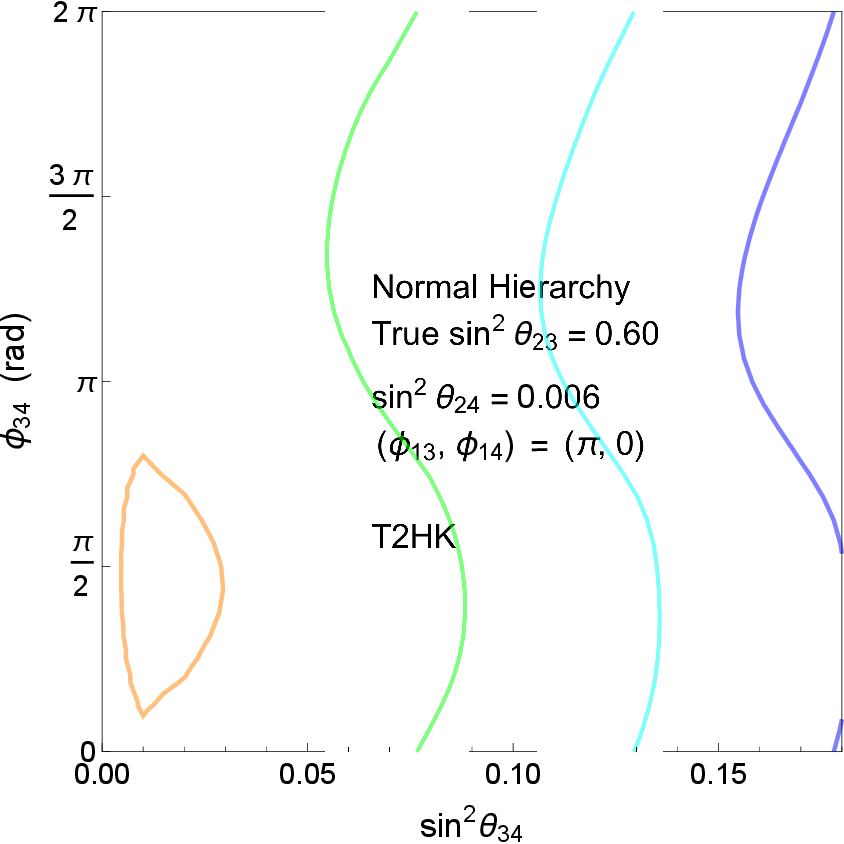}
    \includegraphics[width=55mm]{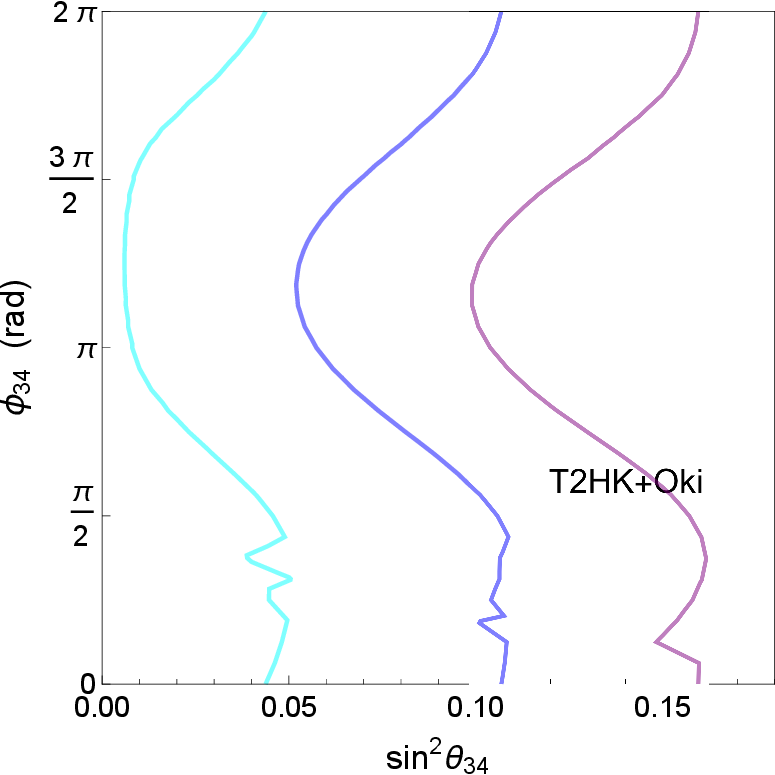}
    \includegraphics[width=55mm]{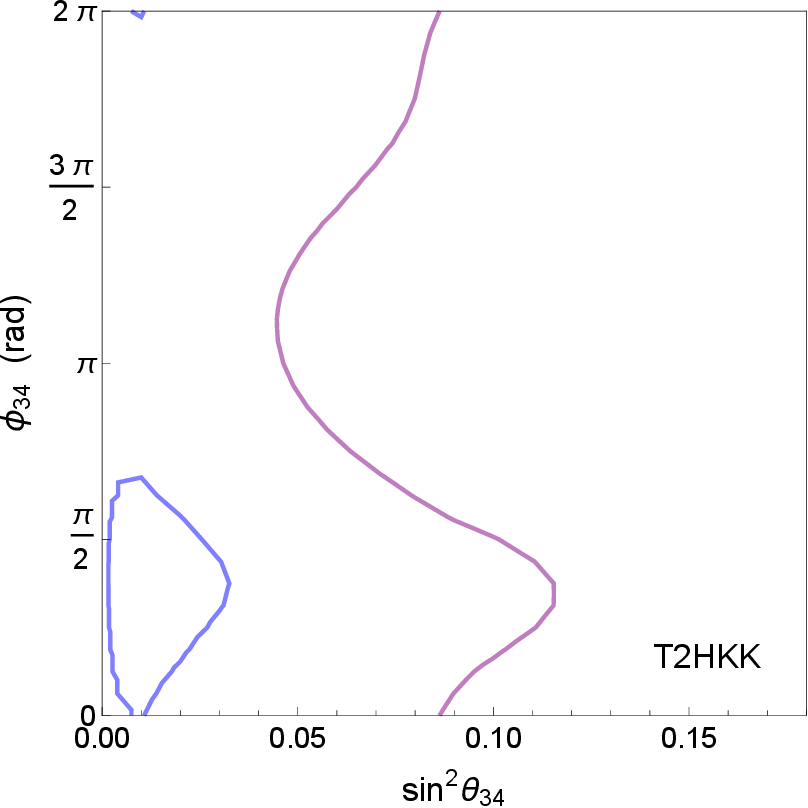}
    \caption{
    Significance of rejecting the wrong $\theta_{23}$ octant when the true $\sin^2\theta_{23}$ is 0.44,
     Eq.~(\ref{chisq044}) (upper three panels), and that when the true $\sin^2\theta_{23}$ is 0.60,
     Eq.~(\ref{chisq060}) (lower three panels),
     in T2HK, T2HKK and T2HK+Oki experiments, in the second benchmark Table~\ref{physpara2}.
     The mass hierarchy is normal.
    The orange, green, cyan, blue and purple contours correspond to
 $\Delta \chi^2=1.5^2,\ 2^2,\ 2.5^2,\ 3^2, \ 3.5^2$, respectively.
    }
    \label{nh34}
  \end{center}
\end{figure}

\begin{figure}[H]
  \begin{center}
    \includegraphics[width=55mm]{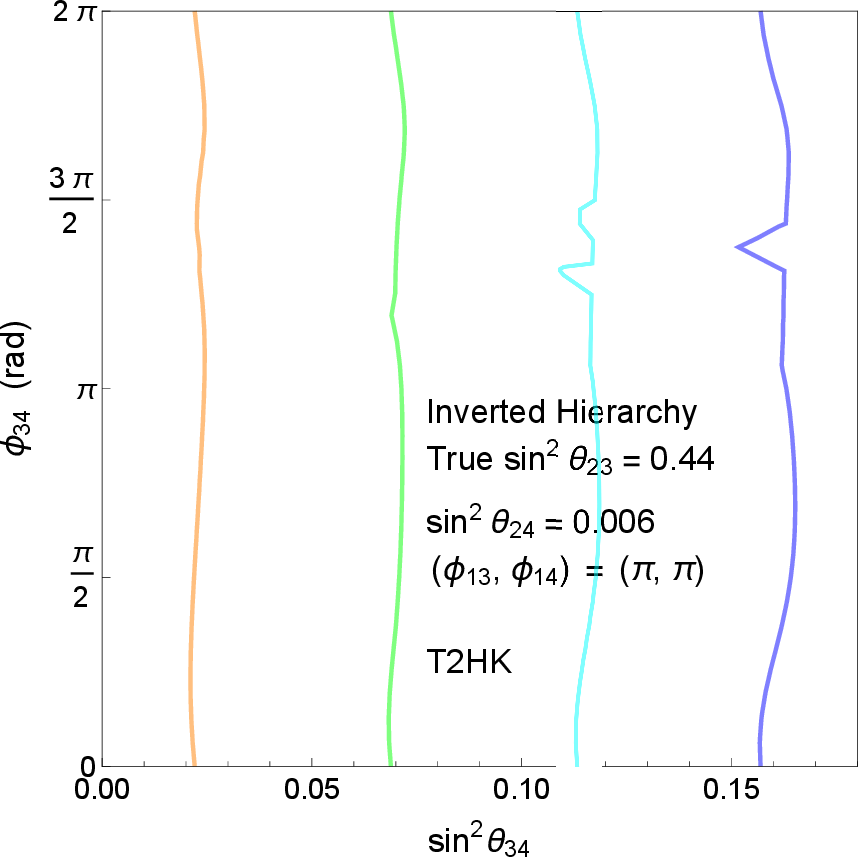}
    \includegraphics[width=55mm]{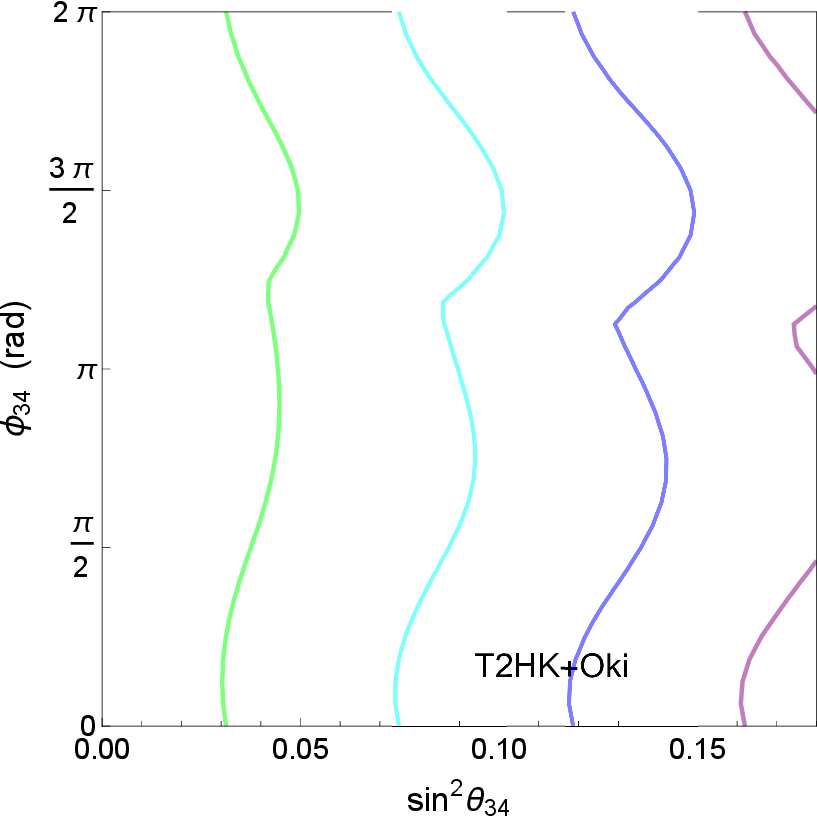}
    \includegraphics[width=55mm]{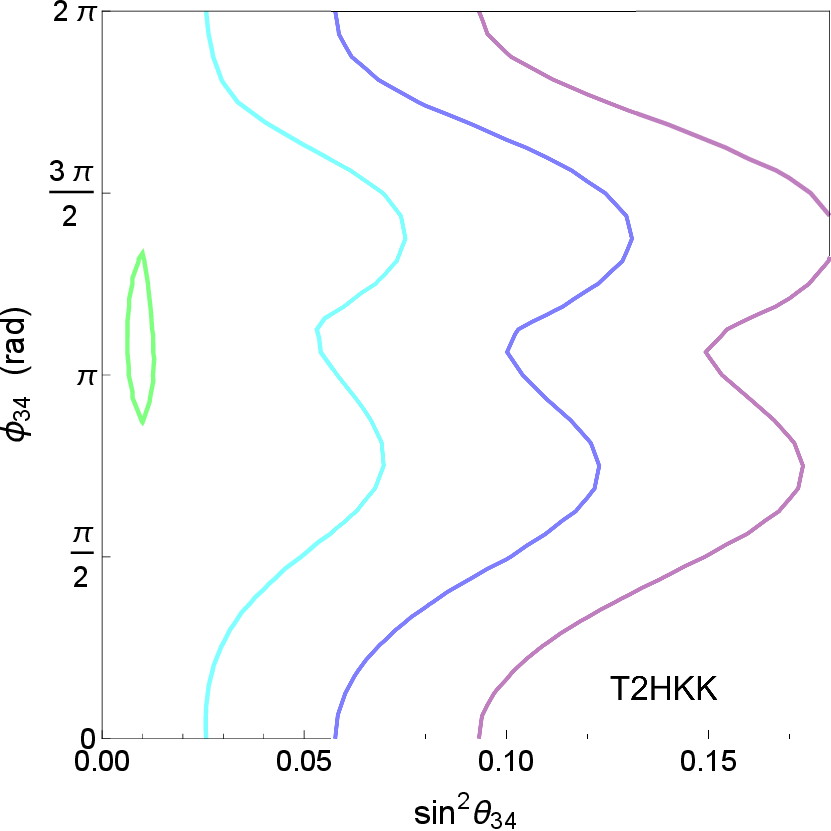}
    \\
        \includegraphics[width=55mm]{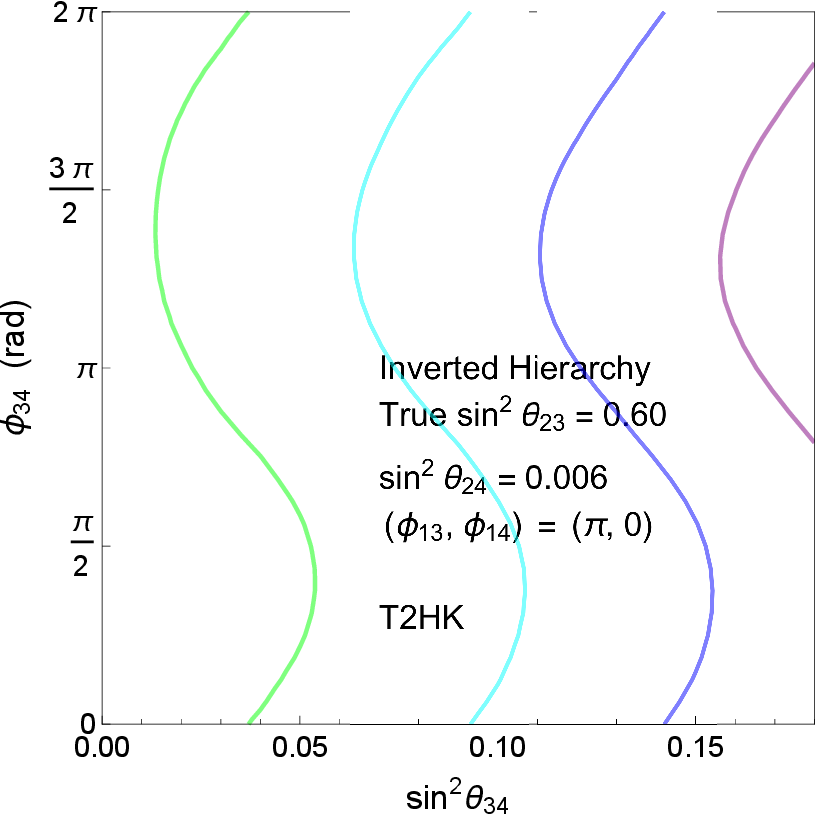}
    \includegraphics[width=55mm]{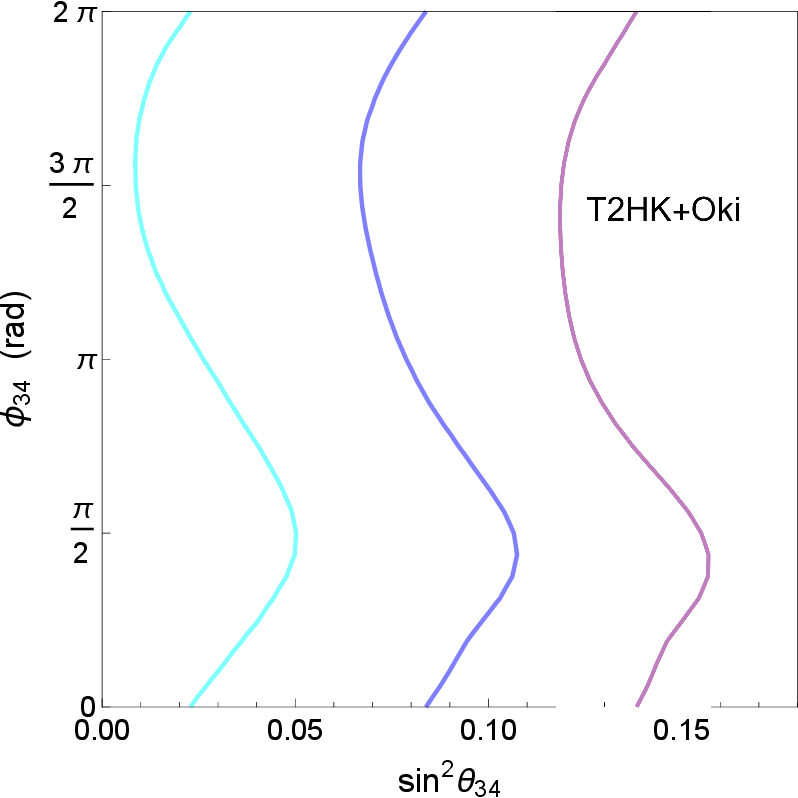}
    \includegraphics[width=55mm]{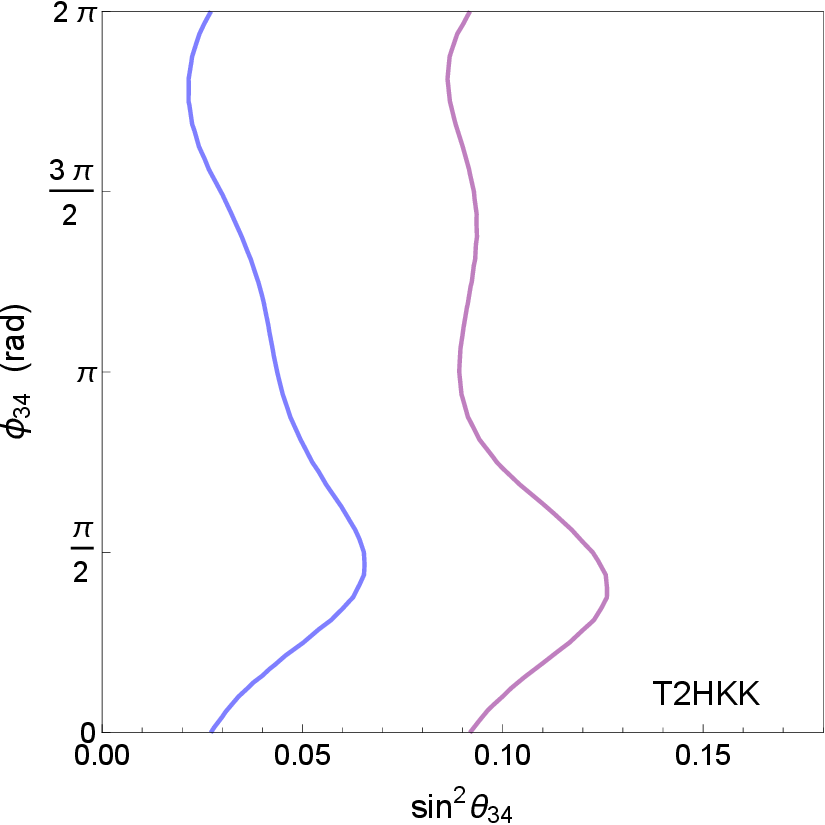}
    \caption{
    Same as Fig.~\ref{nh34} except that the mass hierarchy is inverted.
    }
    \label{ih34}
  \end{center}
\end{figure}

Next, we present $\Delta \chi^2$ with $\sin^2\theta_{23}=$0.44 Eq.~(\ref{chisq044})
 and that with $\sin^2\theta_{23}=$0.60 Eq.~(\ref{chisq060}) in the second benchmark Table~\ref{physpara2}.
The results are presented as a function of $(\sin^2\theta_{34},\,\phi_{34})$ 
 and shown separately for the normal and inverted mass hierarchies.
Fig.~\ref{nh34} is the result when the mass hierarchy is normal,
 and Fig.~\ref{ih34} is the result when the mass hierarchy is inverted.

The following observations are made from Figs.~\ref{nh34},\ref{ih34}.

\begin{itemize}

\item
T2HKK and T2HK+Oki give a larger significance of rejecting the wrong $\theta_{23}$ octant than T2HK 
 in all cases.
Also, T2HKK shows a larger significance than T2HK+Oki in all cases.

\item
Non-zero values of $\theta_{34}$ tend to increase the significance of rejecting the wrong $\theta_{23}$ octant
 in both cases with $\sin^2\theta_{23}=0.44$ and $0.60$, for both mass hierarchies, in all the experiments.
Interestingly, even though the matter effect is small in T2HK, we still obtain an increase in the significance in T2HK.
The above results are because the term Eq.~(\ref{matter2}) is proportional to $\sin\theta_{23}$ and flips sign
 for neutrinos and antineutrinos.
Hence, a measurement of this term using the combination of neutrino-focusing and antineutrino-focusing operations
 offers a new probe for the $\theta_{23}$ octant.

\item
In the case with $\sin^2\theta_{23}=0.60$ and for small $\sin^2\theta_{34}$,
 there is a tiny parameter region where the significance of rejecting the wrong $\theta_{23}$ octant decreases with $\theta_{34}$.
The presence of such a region is probably because the term Eq.~(\ref{matter2}) mimics the first part of Eq.~(\ref{sinsq})
 (remember that $\phi_{13}=\pi$ in the second benchmark and so the true value of the first part of Eq.~(\ref{sinsq}) is 0),
 which makes it harder to distinguish the term Eq.~(\ref{sinsq}) from the term Eq.~(\ref{main}).

\end{itemize}

\begin{figure}[H]
  \begin{center}
    \includegraphics[width=55mm]{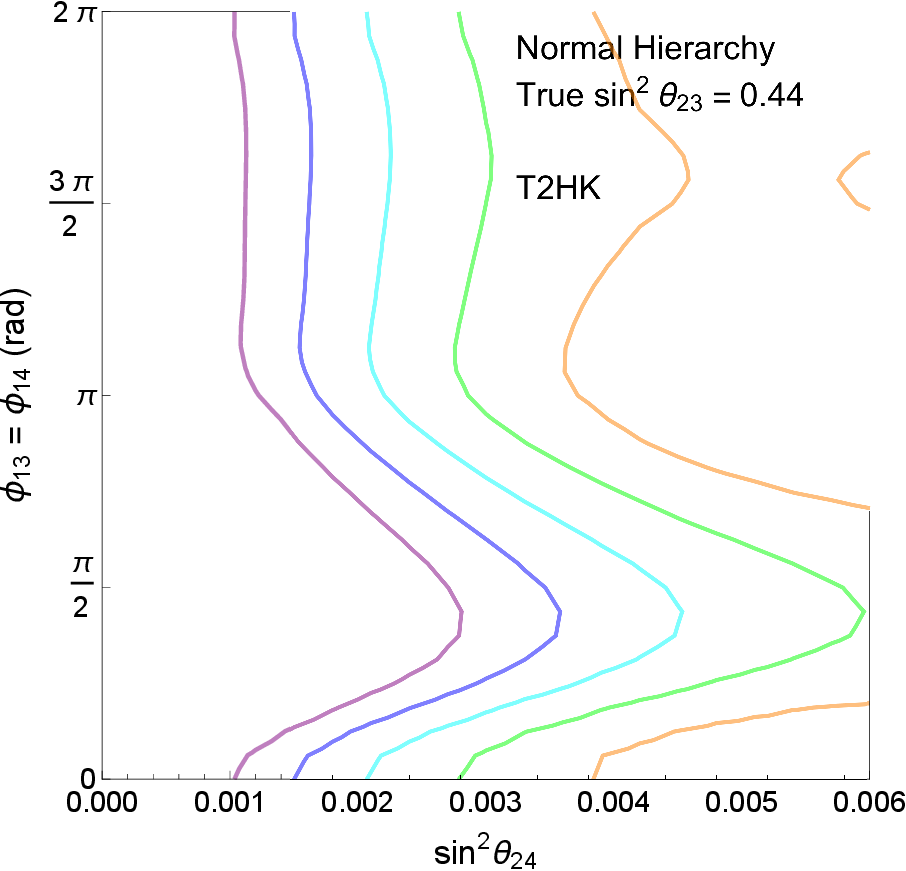}
    \includegraphics[width=55mm]{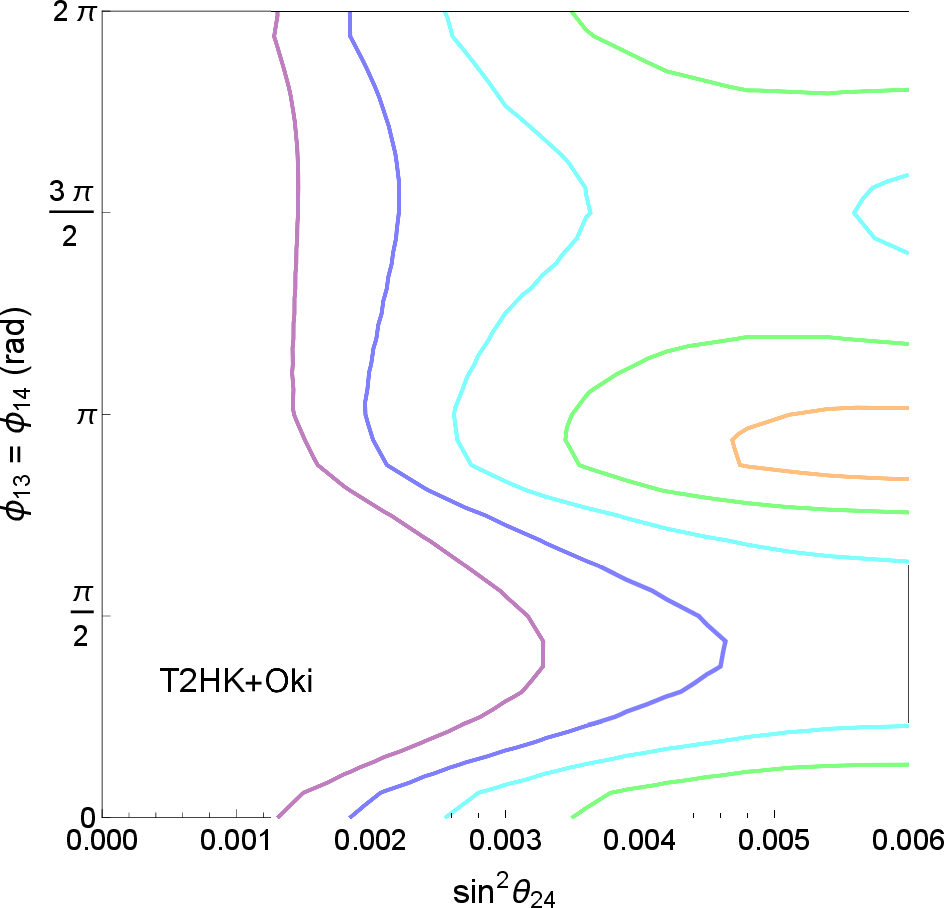}
    \includegraphics[width=55mm]{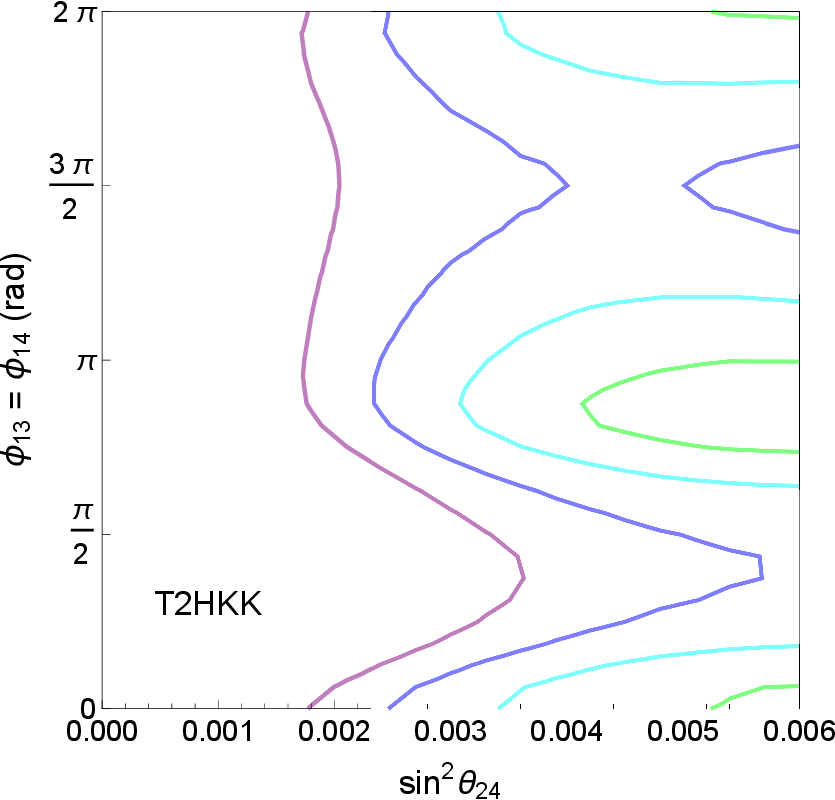}
    \\
        \includegraphics[width=55mm]{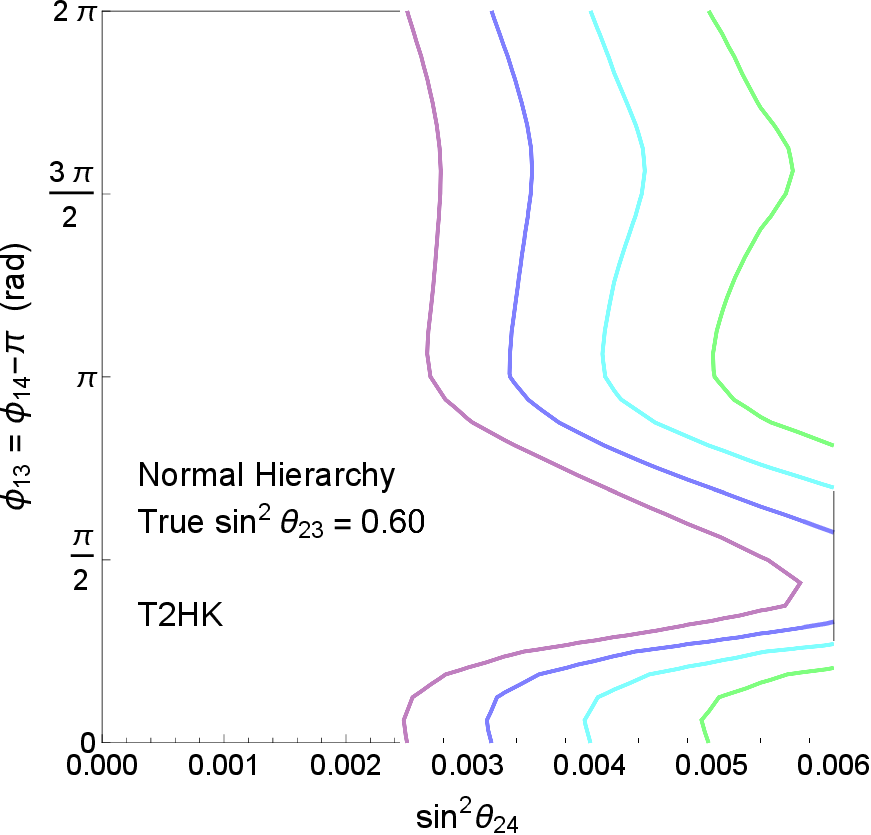}
    \includegraphics[width=55mm]{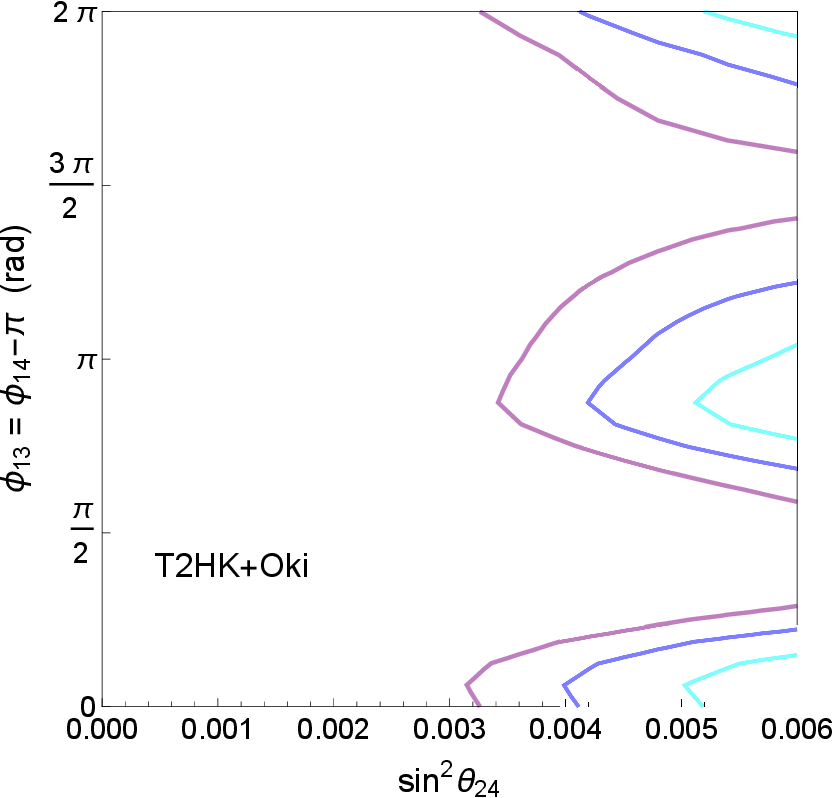}
    \includegraphics[width=55mm]{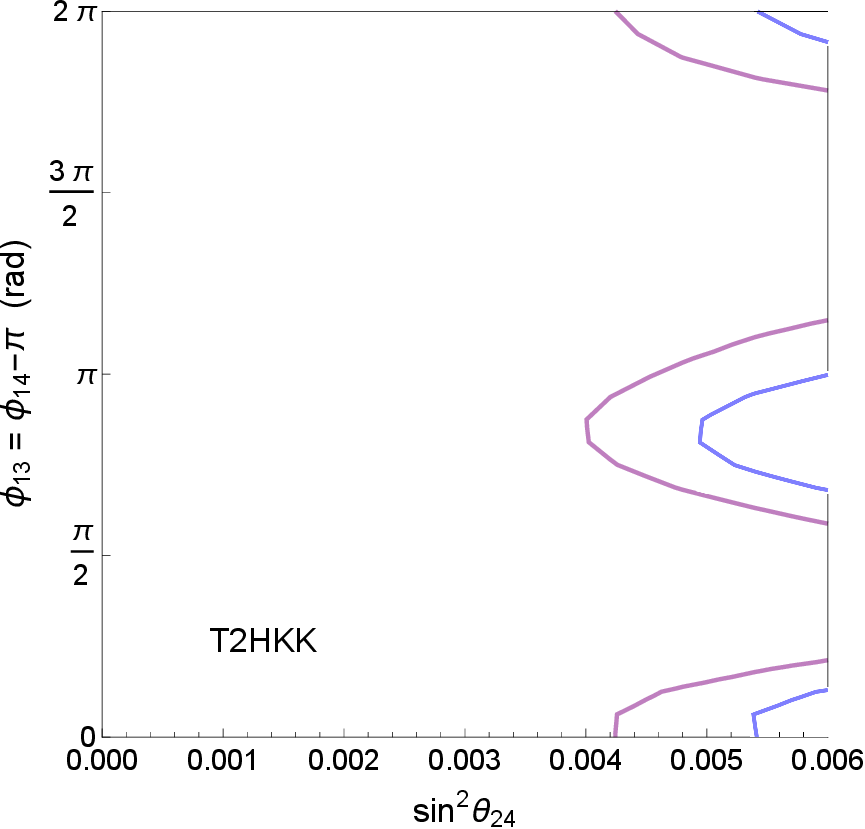}
    \caption{
    Significance of rejecting the wrong $\theta_{23}$ octant when the true $\sin^2\theta_{23}$ is 0.44,
     Eq.~(\ref{chisq044}) (upper three panels), and that when the true $\sin^2\theta_{23}$ is 0.60,
     Eq.~(\ref{chisq060}) (lower three panels), in T2HK, T2HKK and T2HK+Oki experiments, in the third benchmark Table~\ref{physpara3}.
     The mass hierarchy is normal.
    The orange, green, cyan, blue and purple contours correspond to
 $\Delta \chi^2=1.5^2,\ 2^2,\ 2.5^2,\ 3^2,\ 3.5^2$, respectively.
    }
    \label{nh24}
  \end{center}
\end{figure}

\begin{figure}[H]
  \begin{center}
    \includegraphics[width=55mm]{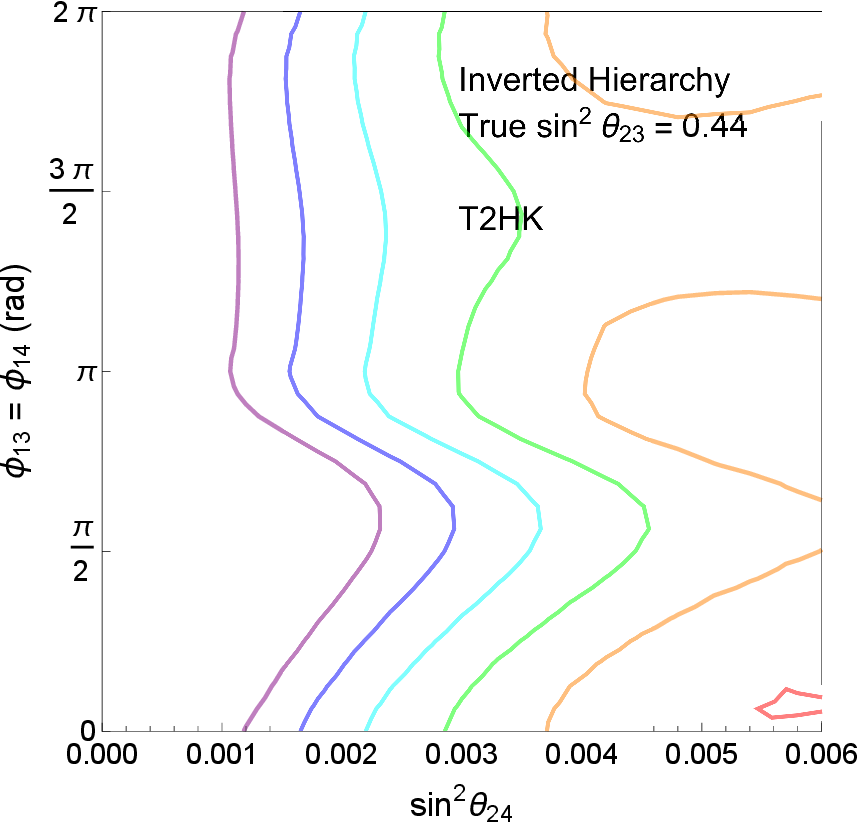}
    \includegraphics[width=55mm]{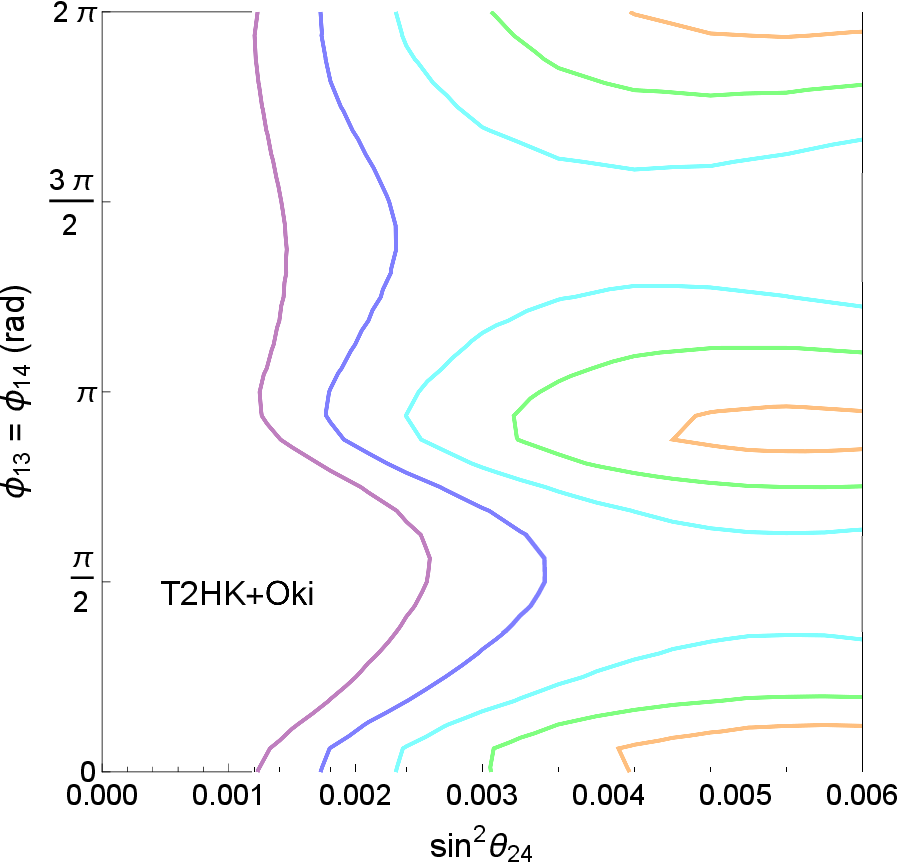}
    \includegraphics[width=55mm]{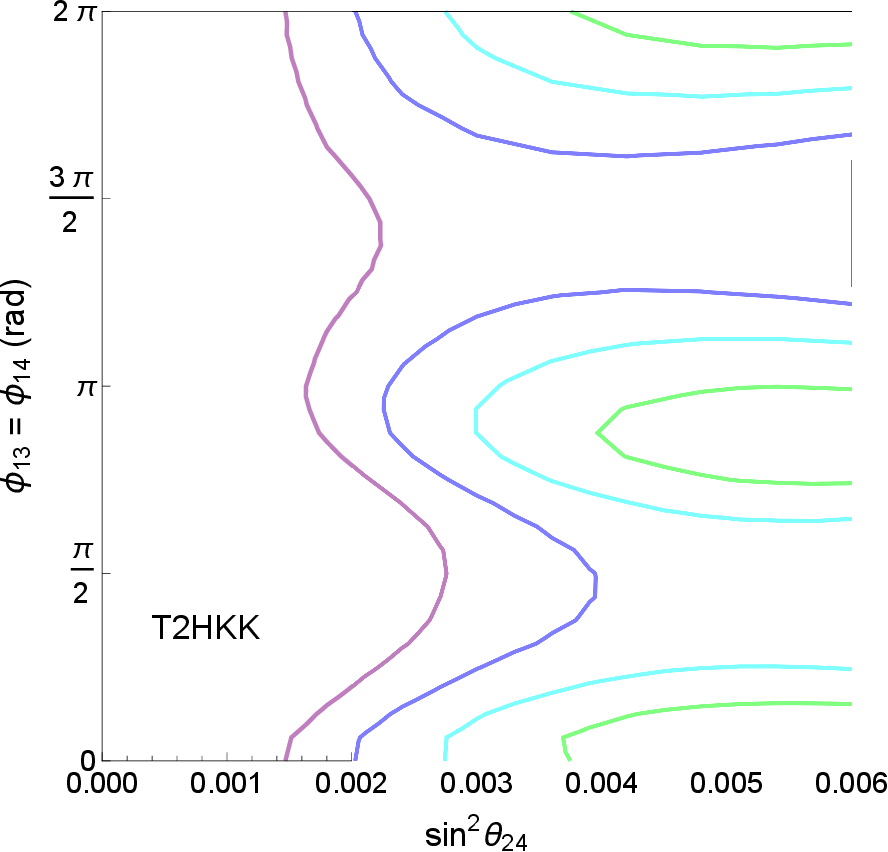}
    \\
        \includegraphics[width=55mm]{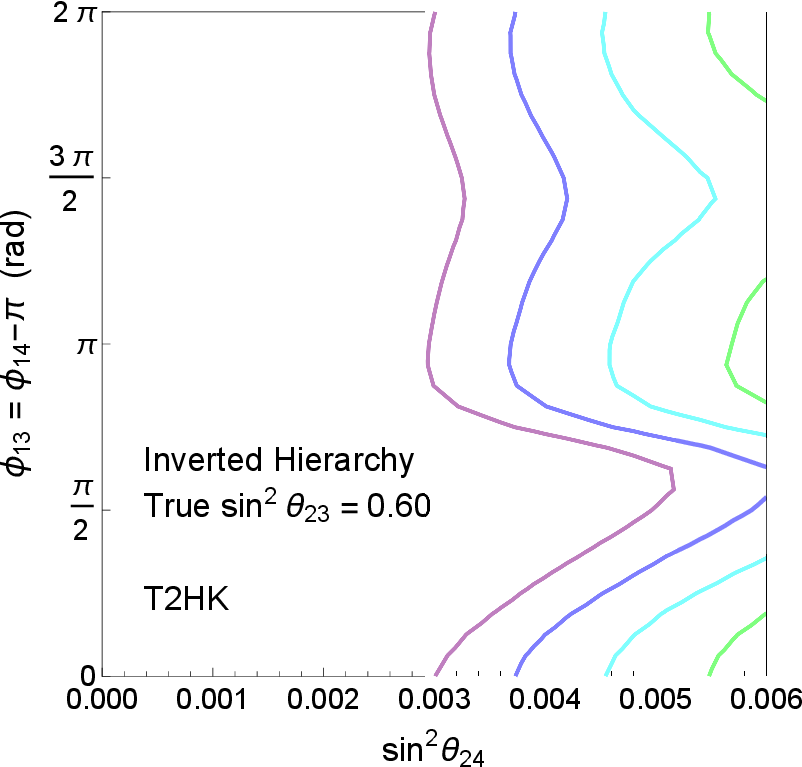}
    \includegraphics[width=55mm]{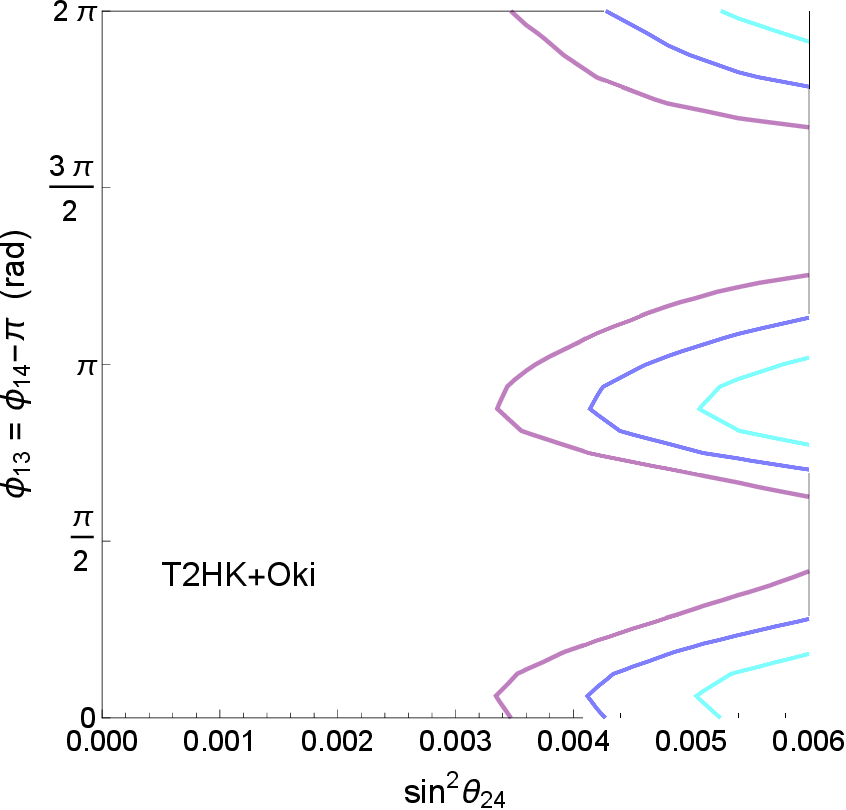}
    \includegraphics[width=55mm]{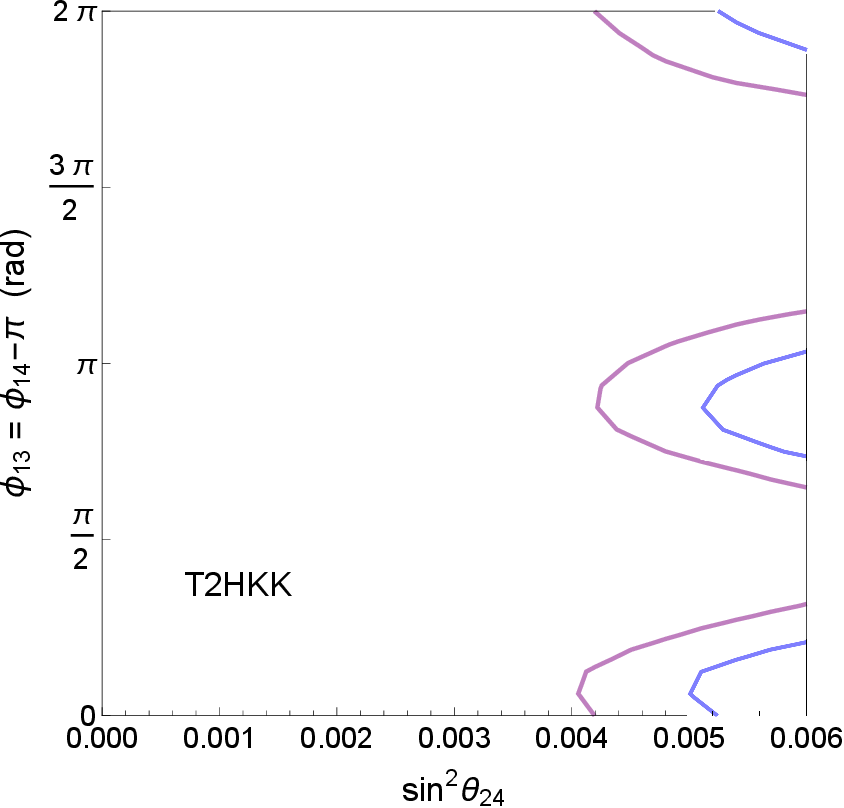}
    \caption{
    Same as Fig.~\ref{nh24} except that the mass hierarchy is inverted. The red contour corresponds to $\Delta \chi^2=1$.
    }
    \label{ih24}
  \end{center}
\end{figure}

Finally, we present $\Delta \chi^2$ with $\sin^2\theta_{23}=$0.44 Eq.~(\ref{chisq044})
 and that with $\sin^2\theta_{23}=$0.60 Eq.~(\ref{chisq060}) in the third benchmark Table~\ref{physpara3}.
The results are presented as a function of $(\sin^2\theta_{24},\,\phi_{13}=\phi_{14})$ for $\sin^2\theta_{23}=$0.44
 and $(\sin^2\theta_{24},\,\phi_{13}=\phi_{14}-\pi)$ for $\sin^2\theta_{23}=$0.60,
 and shown separately for the normal and inverted mass hierarchies.
Fig.~\ref{nh24} is the result when the mass hierarchy is normal,
 and Fig.~\ref{ih24} is the result when the mass hierarchy is inverted.

The following observations are made from Figs.~\ref{nh24},\ref{ih24}.

\begin{itemize}

\item
T2HKK and T2HK+Oki give a larger significance of rejecting the wrong $\theta_{23}$ octant than T2HK 
 in all cases.
Also, T2HKK shows a larger significance than T2HK+Oki in all cases.

\item
The significance of rejecting the wrong $\theta_{23}$ octant mostly decreases with $\sin^2\theta_{24}$
 in both cases with $\sin^2\theta_{23}=0.44$ and $0.60$, for both mass hierarchies, in all the experiments.
The significance in T2HK is above 3 for $\sin^2\theta_{24}<0.0015$ (with $\sin^2\theta_{14}=0.004$)
 when $\sin^2\theta_{23}=0.44$,
 and for $\sin^2\theta_{24}<0.003$ (with $\sin^2\theta_{14}=0.004$) when $\sin^2\theta_{23}=0.60$.
This suggests that the sensitivity of T2HK to the $\theta_{23}$ octant is recovered for such small values of sterile-active mixings.

\item
When $\sin^2\theta_{23}=0.44$, for $\phi_{13}\simeq3\pi/2$ and for large $\sin^2\theta_{24}$,
 there is a tiny parameter region where the significance increases with $\sin^2\theta_{24}$.
The presence of such a region is probably because for larger $\sin^2\theta_{24}$,
 the first part of Eq.~(\ref{sinsq}) is less likely to mimic the second part of Eq.~(\ref{sin2})
 (remember that $\phi_{13}-\phi_{14}=0,\pi$ in the third benchmark and so the true value of the second part of Eq.~(\ref{sin2}) is zero),
 which makes it easier to distinguish the first part of Eq.~(\ref{sinsq}) from the other terms.
Then the combination of neutrino- and antineutrino-focusing operations can more easily 
 resolve the degeneracy between Eq.~(\ref{main}) and Eq.~(\ref{sinsq}).

\end{itemize}

We note in passing that in the $\chi^2$ minimum in the wrong octant region, the fitted value of $\phi_{13}$ is close to its true value.
So, even if the sensitivity to the $\theta_{23}$ octant is lost, the measurement of the standard CP phase is relatively unaffected.
\label{results}

%%%%%%%%%%%%%%%
\section{Summary}

We have confirmed that in the presence of sterile-active mixings with non-zero $\theta_{14}$ and $\theta_{24}$,
 T2HK experiment with two 187~kton detectors at Kamioka 
 can lose sensitivity to the $\theta_{23}$ octant, depending on values of CP phases.
We have revealed that T2HKK exhibits a better sensitivity to the $\theta_{23}$ octant in all parameter regions including those where
 the experiment with two detectors at Kamioka loses its sensitivity, in spite of smaller statistics of T2HKK.
The better sensitivity of T2HKK is because measurements of the same beam at Kamioka and Korea,
 which involve distinctively different baseline lengths, resolve the degeneracy between the atmospheric oscillation probability and the sum of the interferences with the solar oscillation amplitude and active-sterile oscillation amplitude.
We have further studied the impact of non-zero $\theta_{34}$ mixing and found that the sensitivity to the $\theta_{23}$ octant
 tends to increase with $\theta_{34}$ in all parameter regions in all the experiments.

Our results suggest that if a hint of sterile-active mixings with $\theta_{14}\neq0$ and $\theta_{24}\neq0$
 is discovered but $\theta_{34}$ is 0 or substantially smaller than the current experimental bound,
 T2HKK is a preferable option compared to the plan of placing the two detectors at Kamioka,
 for the determination of the $\theta_{23}$ octant.
 
Additionally, we have studied a case where one 187~kton detector is at Kamioka and a 100~kton detector is at Oki Islands.
This case also shows a better sensitivity to the $\theta_{23}$ octant in those parameter regions where the plan of two 187~kton detectors loses its sensitivity, but the improvement is quite mild compared to T2HKK, in spite of its larger statistics than T2HKK.

%%%%%%%%%%%%%%%
\section*{Acknowledgement}
This work is partially supported by Scientific Grants by the Ministry of Education, Culture, Sports, Science and Technology of Japan,
Nos.~17K05415, 18H04590 and 19H051061 (NH), and No.~19K147101 (TY).

%%%%%%%%%%%%%%%
%%% References %%%
%%%%%%%%%%%%%%%

\end{document}